\documentclass[aps,prstab,twocolumn,groupedaddress,nofootinbib,floatfix]{revtex4-1}

\usepackage{amsmath}
\usepackage{graphicx}
\usepackage[caption=false]{subfig}
\usepackage{color}

\begin{document}


\title{Intrabeam Scattering Studies at the Cornell Electron-positron Storage Ring Test Accelerator}


\author{M. P. Ehrlichman}
\email[]{mpe5@cornell.edu}
\author{W. Hartung}
\author{B. Heltsley}
\author{D. P. Peterson}
\author{N. Rider}
\author{D. Rubin}
\author{D. Sagan}
\author{J. Shanks}
\author{S. T. Wang}
\affiliation{Cornell Laboratory for Accelerator-based Sciences and 
Education, Cornell University, Ithaca, NY}
\author{R. Campbell}
\author{R. Holtzapple}
\affiliation{Department of Physics, 
California Polytechnic State University, San Luis Obispo, CA}


\date{\today}

\begin{abstract}
Intrabeam scattering (IBS) limits the emittance and single-bunch 
current that can be achieved in electron or positron storage ring 
colliders, damping rings, and light sources.  Much theoretical 
work on IBS exists, and while the theories have been validated
in hadron and ion machines, the presence of strong damping 
makes IBS in lepton machines a different phenomenon.  We present 
the results of measurements at CesrTA of IBS dominated beams, 
and compare the data with theory.  The beams we study have 
parameters typical of those specified for the next generation 
of wiggler dominated storage rings: low emittance, small bunch 
length, and an energy of a few GeV.  Our measurements are in good 
agreement with IBS theory, provided a tail-cut procedure is applied.

\end{abstract}


\maketitle

\section{Introduction\label{sec:introduction}}
Next-generation lepton storage rings are presently being designed for
light sources, damping rings, and other applications 
\cite{ilc:tr,clic:ipac12,prstab:usr,erhic:report}.
These designs are intended to reach new records for high stored
currents and low emittances.  This will require new
accelerators to operate with higher charge per bunch, more bunches
per beam, and smaller bunch dimensions.
Intrabeam scattering (IBS) is a single-bunch, collective
effect that limits the density of particle beams \cite{kmw:2005}
which will likely be one of the mechanisms that limit the performance
of future rings.  The consequences of IBS can be interpreted as 
either a per-bunch current limit or a lower bound on the emittance of a 
bunch with a given charge.  These limits
depend on the optics, beam energy, radiation damping time, etc.

Intrabeam scattering has been studied in detail at $p$ and $\bar{p}$ 
\cite{lebedev:2005,martini:1984,piwinski:desy}, 
and heavy ion colliding beam machines \cite{rhic-exp:2006}.
In such machines, IBS slowly dilutes the emittance of the beam and 
imposes a luminosity lifetime.
Good agreement was found between IBS theory
and experiment at the Relativistic Heavy Ion 
Collider (RHIC) at Brookhaven National Lab \cite{rhic-exp:2006}.
Lattices which reduce IBS growth by minimizing the dispersion invariant
${\cal H}_a=\gamma_a\eta_a^2+2\alpha_a\eta_a\eta'_a+\beta_a\eta'^2_a$ 
have been implemented at RHIC and are used 
regularly for colliding-beam experiments \cite{rhic-supp:2008}.  
For beams of protons and anti-protons,
good agreement between theory and
measurements was found at the Tevatron \cite{lebedev:2005}.

Electron and positron beams in rings reach equilibrium much more rapidly
than hadron beams, hence IBS in lepton rings manifests itself
differently.
Lepton machines have strong radiation
damping, and the equilibrium emittance is determined by a balance between 
radiation damping and quantum excitation.  
Typical damping times are on the order of tens of milliseconds.
The quantized nature of IBS contributes a random motion to the scattered 
particles, which tends to increase the 
phase space volume of the bunch. The random excitation due to IBS 
equilibrates with radiation damping to determine 
the beam size. The result is a current-dependent emittance.

Large-angle scattering events that kick particles outside the core of the
bunch and contribute to particle loss and beam halo are relatively rare.
Small-angle scattering events are more common.
The former are commonly
referred to as Touschek scattering, and the latter as intrabeam scattering. 
The emphasis in this paper is on intrabeam scattering.

IBS in electron beams has been studied at the Accelerator Test Facility 
(ATF) at KEK \cite{atf:2002}, where
detailed measurements of the current dependence of the bunch energy 
spread and length are in good agreement with theory.  Measurements
of the transverse dimensions at ATF, however, are not as complete.  

CesrTA is a re-purposing of the Cornell Electron Storage Ring (CESR) 
as a test accelerator for future
low-emittance storage rings designs \cite{cesrta:2009}.
CesrTA is a wiggler-dominated storage ring, with 90\% of the synchrotron 
radiation produced by twelve $1.9$~T superconducting damping wigglers.
Some parameters for CesrTA are given in Table \ref{tab:cesr-params}.
Design and analysis of CesrTA is done using the {\tt Bmad} relativistic
charged beam simulation library \cite{bmad:2006}.
Measured $a$-mode (horizontal-like), single particle geometric 
emittance $\epsilon_a$ is $3.4$~nm-rad.  
The minimum measured $b$-mode (vertical-like) emittance at the time of 
these measurements is $\epsilon_b \approx 20$~pm-rad, and arises from 
sources such as magnet alignment, field
errors, and quality of beam-based optics corrections. Subsequent machine 
studies have reduced the $b$-mode emittance by another 50\%, at which 
point the $b$-mode emittance is
dominated by sources unaffected by optics correction \cite{prstab:let}.
The flexibility of the CesrTA optics allows precise control of $b$-mode 
emittance above that minimum.  We are able to vary $b$-mode emittance
by using closed coupling bumps to introduce a localized vertical 
dispersion in the damping wigglers. In this way,
vertical emittance can be increased by an order of magnitude without 
affecting the global optics.  The bunch length 
is determined by the RF accelerating 
voltage.  With a voltage of $6$~MV, the bunch length is about $10.5$~mm.
Measurements were made with bunch charges ranging from 
$1.6\times 10^9$ to $1.6\times 10^{11}$ particles/bunch 
($0.1$~mA to $10$~mA).
\begin{table}[tb]
\centering
\caption{Machine parameters for IBS measurements.\label{tab:cesr-params}}
\begin{tabular*}{\columnwidth}{@{\extracolsep{\fill}}lr}
\hline
\hline
Beam Energy (GeV)                          &$2.085$ \\
Circumference (m)                          &$768$ \\
RF Frequency (MHz)                         &$499.765$ \\
Horizontal Tune ($Q_x$)                    &$14.624$ \\
Vertical Tune ($Q_y$)                      &$9.590$ \\
Synchrotron Tune ($Q_z$)                   &$-0.065$ \\
Transverse Damping Time (ms)               &$56.6$ \\
\hline
\hline
\end{tabular*}
\end{table}

CesrTA is instrumented for precision bunch size measurements
in all three dimensions.  
Vertical beam size measurements are made using an x-ray beam size 
monitor (xBSM), which images x-rays from a hard bend magnet through 
a pinhole onto a vertical diode detector array 
\cite{xbsm:handbook,rider:2012}.  The instrument images the beam turn-by-turn,
allowing bunch position and size measurement on each bunch passage.
These turn-by-turn images can be analyzed collectively to
reveal beam motion and beam size fluctuations.  The images can also be 
summed over all turns to improve average beam size accuracy at low 
current, after correcting for beam motion.
Horizontal beam size measurements are made 
with a visible-light interferometer \cite{suntao:2012}.
The interferometer is used to image visible synchrotron radiation 
on a charge-coupled device (CCD)
that is exposed for about $400$ turns at high current or about $40000$ 
turns at low current.
Bunch length measurements are done with a streak camera using visible light
from a bending magnet \cite{holtzapple:2000}.  
The horizontal, vertical, and longitudinal data plotted in 
this paper are the binned average over measurements within a current
range.  The error bars are
the statistical uncertainty of the measurements within the bin.

Validation of the beam size instrumentation includes checking for intensity 
dependent systematics using filters, and size systematics by varying 
source-point betatron-functions.  The horizontal beam size monitor  
also undergoes direct calibration 
with a source of known size \cite{suntao:2012}.

One of the goals of the CesrTA IBS investigation is to
improve on the ATF results by including detailed measurements of the bunch
charge dependence of the transverse beam sizes.  
In addition to robust instrumentation,
CesrTA has independently powered quadrupoles and the capability to store 
larger single-bunch charges.  This flexibility allows
for measurements at CesrTA in a greater variety of conditions.

We use the IBS formalism developed by Kubo and Oide \cite{kubooide:2001} 
to describe the data.  The formalism is a generalization 
of the Bjorken-Mtingwa description \cite{bjorken-mtingwa:1983} and 
uses an eigen-decomposition of the beam $\mathbf{\Sigma}$-matrix rather than
the traditional Twiss parameters.  This formalism
naturally handles arbitrary coupling among the three beam dimensions.

In this paper, we describe the CesrTA IBS experiments, 
and compare the results to both analytic theory and Monte 
Carlo simulations.  Some of the results shown here 
were first presented at the 2012 International Particle Accelerator
Conference \cite{ipac-ibs:2012}.  The present paper provides 
a more complete description and theoretical framework
for the results.  Further details can be found in \cite{ehrlichman:2013}.

\section{Theory\label{sec:theory}}
The IBS formalism outlined here is described succinctly by Kubo 
\cite{sad:1999} and in detail by Kubo and Oide \cite{kubooide:2001}.
It is based on changes to the second-order moments 
of the $\mathbf{\Sigma}$-matrix 
of the beam distribution in the frame of the bunch
\begin{equation}
\Delta\left<\bar{p}_i \bar{p}_j\right> = 
c_I \mathbf{R} \left<\mathbf{\delta w}^2\right> \mathbf{R}^T,
\end{equation}
where $\mathbf{R}$ is a matrix of eigenvectors defined below,
$c_I$ is proportional to the bunch charge, and
\begin{equation}
\left<\mathbf{\delta w}^2\right>=\begin{pmatrix}
\left<\delta w_1^2\right> &0 &0\\
0& \left<\delta w_2^2\right> &0\\
0& 0& \left<\delta w_3^2\right>
\end{pmatrix},
\end{equation}
$\left<\delta w_1^2\right>$, $\left<\delta w_2^2\right>$, and
$\left<\delta w_3^2\right>$ are the rates of change of the 
normal mode 2nd order moments.

\begin{figure}
\includegraphics[width=\columnwidth]{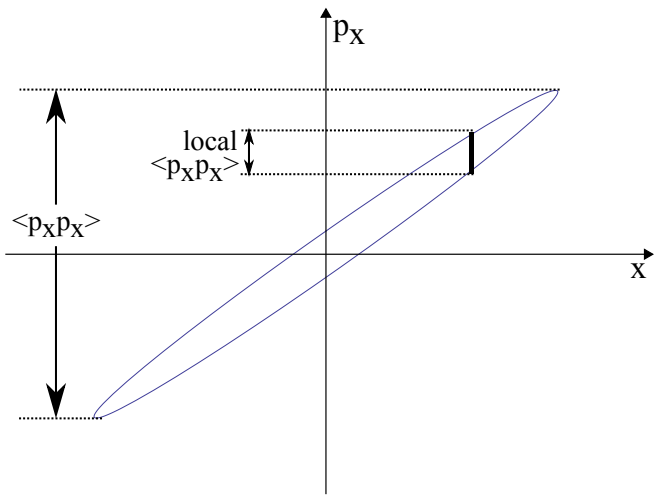}
\caption{The local momentum $\mathbf{\Sigma}$-matrix describes the 
distribution of momentum in a small spatial element of the bunch.
\label{fig:local-momentum-fig}}
\end{figure}

IBS refers to scattering among nearby particles.
The 2nd order moments of the $\mathbf{\Sigma}$-matrix describe the momentum 
spread 
of the entire bunch.  What is needed is the ``local'' momentum spread, or the 
spread in the momentum of particles inside a small spatial element 
of the bunch.
The difference between the $\mathbf{\Sigma}$-matrix 2nd order moments 
and the ``local'' moments is depicted in Fig.~\ref{fig:local-momentum-fig}.  
The local momentum spread is obtained via
\begin{equation}
\mathbf{\Sigma}_{lpp}\equiv \left<\bar{p}_{li}\bar{p}_{lj}\right> 
= \mathbf{\Sigma}_{pp} - \mathbf{\Sigma}_{xp}^T \mathbf{\Sigma}_{xx}^{-1} 
\mathbf{\Sigma}_{xp}\textrm{,}
\label{eqn:local-momentum-spread}
\end{equation}
where $\mathbf{\Sigma}_{pp}\equiv\left<\bar{p}_i \bar{p}_j\right>$, 
$\mathbf{\Sigma}_{xx}\equiv\left<\bar{x}_i \bar{x}_j\right>$, and
$\mathbf{\Sigma}_{xp}\equiv\left<\bar{x}_i \bar{p}_j\right>$.

$\mathbf{\Sigma}_{lpp}$ is symmetric and 
positive-definite and can be decomposed as
\begin{equation}
\mathbf{\Sigma}_{lpp} = \mathbf{R} \mathbf{G} \mathbf{R}^T\textrm{,}
\end{equation}
where $\mathbf{G}$ is a diagonal matrix of the eigenvalues of 
$\mathbf{\Sigma}_{lpp}$ and the 
columns of $\mathbf{R}$ are the eigenvectors.  
The eigenvalues are denoted $u_1$, $u_2$, $u_3$.  
Note that $\mathbf{R}^T=\mathbf{R}^{-1}$.

$\left<\mathbf{\delta w}^2\right>$ is obtained from
\begin{eqnarray}
\left<\delta w_1^2\right>&=&g_2+g_3-2g_1\textrm{,}\label{eqn:w1}\\
\left<\delta w_2^2\right>&=&g_1+g_3-2g_2\textrm{,}\label{eqn:w2}\\
\left<\delta w_3^2\right>&=&g_1+g_2-2g_3\textrm{,}\label{eqn:w3}
\end{eqnarray}
where
\begin{eqnarray}
g_1&=&g\left(u_1,u_2,u_3\right)\textrm{,}\\
g_2&=&g\left(u_2,u_1,u_3\right)\textrm{,}\\
g_3&=&g\left(u_3,u_1,u_2\right)\textrm{,}
\end{eqnarray}
and
\begin{multline}
g\left(a,b,c\right)=\\
\int_0^{\pi/2}\frac{2 a \sin^2 s \cos s}
{\sqrt{\left(\sin^2 s + \frac{a}{b}\cos^2 s\right)\left(\sin^2 s + 
\frac{a}{c}\cos^2 s\right)}}ds\textrm{.}
\end{multline}
$g_1$, $g_2$, and $g_3$ are analogous to the temperatures of the 
three normal modes of the bunch.

$c_I$ is defined as
\begin{equation}
c_I=\frac{r_e^2 N_e \Delta s}{4\pi\gamma^4\epsilon_a\epsilon_b\epsilon_c}
C_{\Lambda}\textrm{,}
\end{equation}
where $\epsilon_a$, $\epsilon_b$, and $\epsilon_c$ are the normal mode 
emittances of the beam, and the Coulomb logarithm 
$C_{\Lambda}$ will be defined in the next section.  $N_e$ is the 
number of particles in the bunch, $r_e$ is the 
classical electron radius, $\gamma$ is the relativistic factor, 
and $\Delta s$ is the length of the region over which particles
interact.

\subsection{Coulomb logarithm}
The Coulomb log, $C_{\Lambda}$, appears in the integration of the 
Rutherford scattering cross-section over all scattering angles.  
The integral diverges for small scattering angles, which correspond 
to large impact parameters.  This requires the introduction 
of a largest impact parameter cutoff.
We follow the prescription by Kubo and Oide \cite{kubooide:2001} and use
the smaller of the mean inter-particle
distance and the smallest beam dimension as the maximum
impact parameter,
\begin{equation}
b_{max}=\min\left(n^{-1/3},\sigma_x,\sigma_y,\gamma\sigma_z\right),
\end{equation}
where $n$ is the particle density in the bunch frame,
\begin{equation}
n=\frac{N_e}{\left(4\pi\right)^{3/2}\sigma_x\sigma_y\gamma\sigma_z}.
\end{equation}

As for the largest scattering angle (smallest impact parameter), both 
Piwinski and Bjorken-Mtingwa assume that $\theta_{max} = \pi/2$.
It was suggested in \cite{raubenheimer:1994} that scattering events 
which occur less frequently than once per particle per radiation
damping time should be excluded from the calculation of the IBS rise time.
This is because such events do not occur frequently enough for the central 
limit theorem to apply and therefore do not contribute to the Gaussian 
core of the beam. Such infrequent events will generate non-Gaussian tails.
It is the size of the Gaussian core that we can measure, so for comparison 
with the data, we exclude contributions to the tails.

The canonical momentum of a particle in an electron/positron storage ring
is the sum over a history of momentum kicks that occur whenever the particle
radiates a photon.  The photon carries away some transverse momentum, but
the emission event can also increase the transverse momentum of the 
particle if the photon is emitted in a region of finite dispersion.
The overall effect of the emission event on the particle's
momentum depends on the action, angle, and local optics where
a photon is emitted.  Because photon emission is stochastic and occurs
at random points along the particle's trajectory, the kicks
are also stochastic.  According to the central limit
theorem, the momenta of particles in a bunch will be normally distributed
in the absence of IBS.  
In the presence of IBS, the distribution consists of a core which is 
close to Gaussian, along with non-Gaussian tails.

The amount of transverse momentum taken away by the radiated photon tends 
to be larger for particles with a larger transverse momentum.  
This leads to damping and results
in an equilibrium distribution of momenta, rather than unbounded
momentum diffusion.
Perturbations to particle motion damp exponentially with a characteristic 
radiation damping time.  Within one damping time, a large number of 
stochastic photon emission events occur.
For CesrTA, about $20\times10^6$ photons are emitted per 
particle per damping time.

Similarly, there are a large number of small-angle intrabeam scattering 
events that likewise excite oscillations.  
These IBS events increase the width of the momentum distribution.
However, very few large-angle scattering events occur per damping time.

\begin{figure}
\centering
\subfloat{
  \includegraphics[width=0.90\columnwidth]{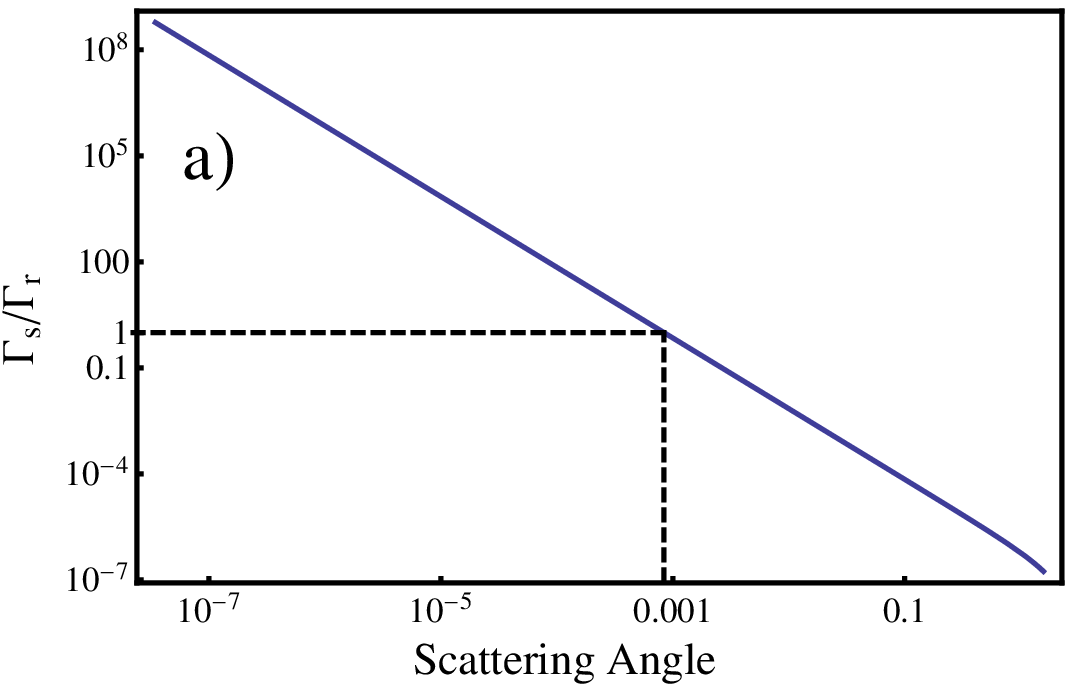}
  \label{fig:tail-cut-rates:a}
}\\
\subfloat{
  \includegraphics[width=0.90\columnwidth]{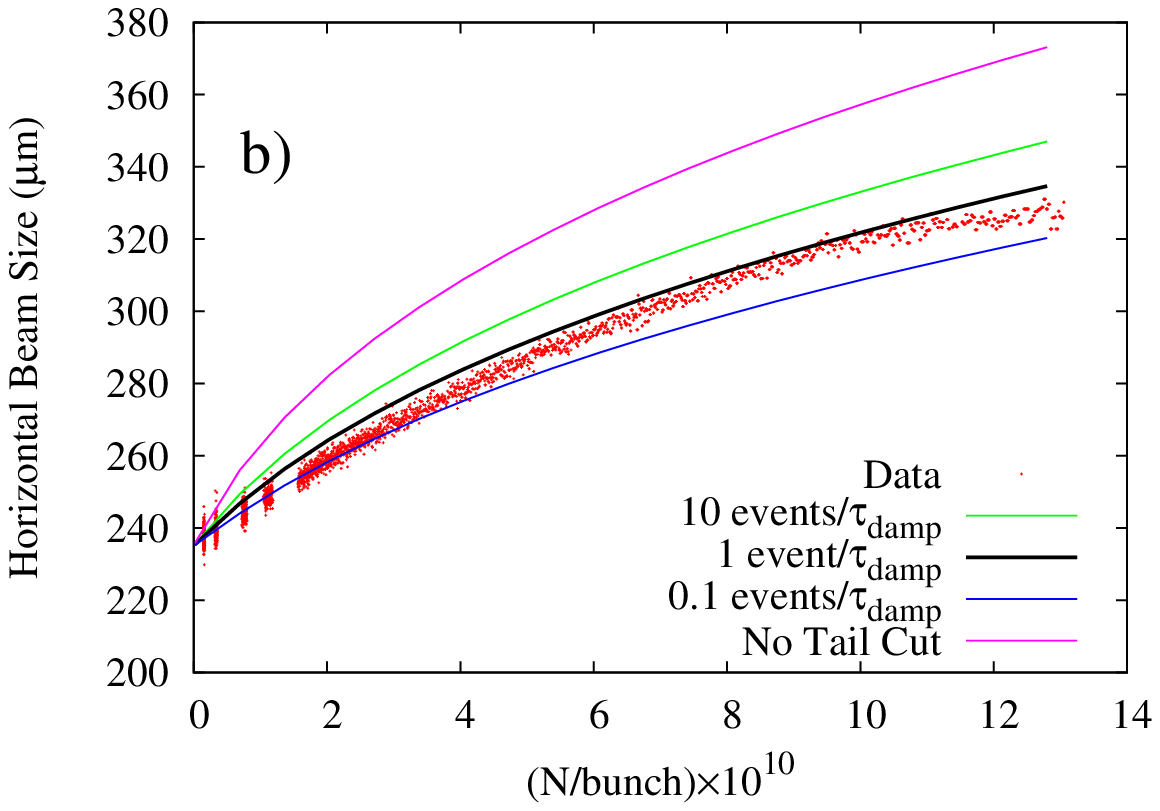}
  \label{fig:tail-cut-rates:b}
}
\caption{\protect\subref{fig:tail-cut-rates:a} Events which occur less than 
once per damping time are excluded from the calculated growth rate.
\protect\subref{fig:tail-cut-rates:b} Equilibrium beam size calculations 
assuming different cut-offs.  \label{fig:tail-cut-rates}}
\end{figure}

A particle with velocity $v$, traveling through a gas with density 
$\rho$, and an interaction cross-section $\sigma$, will undergo scattering 
events at a rate $1/\tau = \rho v\sigma$.  Writing $\sigma=\pi b^2$, 
where $b$ is the effective impact parameter yields
\begin{equation}
\frac{1}{\tau} = \pi\rho v b^2\textrm{.}
\label{eqn:simple-rate-1}
\end{equation}
For non-relativistic Coulomb scattering, the impact parameter is related 
to the scattering angle $\psi$ by
\begin{equation}
b=\frac{r_e}{2\bar{\beta}^2}\cot\frac{\psi}{2}\label{eqn:impact-p}
\end{equation}
where $\bar{\beta}c$ is the velocity of the particles in their 
center-of-momentum frame.  Substituting Equation (\ref{eqn:impact-p}) 
into (\ref{eqn:simple-rate-1}) gives the rate in the lab frame
at which particles 
are scattered into angles less than or equal to $\psi$:
\begin{equation}
\frac{1}{\tau}=\frac{1}{\gamma}\frac{\pi\rho c r_e^2}
{4\gamma^3\left(\epsilon\gamma_a\right)^{\frac{3}{2}}}\cot^2\frac{\psi}{2}
\end{equation}
where $\gamma\sqrt{\epsilon \gamma_a}$ has been used for $\bar\beta$, 
$\epsilon$ is the geometric emittance,
and $\gamma_a$ is the $a$-mode Twiss $\gamma$.
The relevant beam parameters for CesrTA are shown in Table \ref{tab:nominal}.
\begin{table}[tb]
\centering
\caption{Nominal conditions for a bunch with $6.4\times10^{10}$ 
particles.\label{tab:nominal}}
\begin{tabular*}{\columnwidth}{@{\extracolsep{\fill}}lr}
\hline
\hline
Beam Energy $\gamma$                 &$4080$ \\
Average Density $\rho$               &$4.2\times 10^{21}$ part/m$^3$\\
Twiss $\gamma_x$                     &$0.51$ m$^{-1}$ \\
Emittance $\epsilon_a$               &$3.0$ nm-rad\\
\hline
\hline
\end{tabular*}
\end{table}
The rate of scattering events, $\Gamma_s$, in units of 
radiation damping time, $\Gamma_r$, 
as a function of maximum scattering angle
is shown in Fig.~\ref{fig:tail-cut-rates:a}.  The tail-cut
excludes those events which occur less than once per 
radiation damping period.  A measure of the sensitivity to the 
cutoff is illustrated in Fig.~\ref{fig:tail-cut-rates:b}.
The calculated equilibrium beam size is shown for a range of 
two orders of magnitude of the cutoff.  The data 
shown is the same as plotted in Fig. \ref{fig:4B-Methods-horiz}.

The tail-cut consists of restricting the calculation of the IBS growth 
rate to include only those events which occur at least once per particle 
per damping period.
Events which occur less frequently than once per damping period generate
lightly populated non-Gaussian tails that do not contribute to the Gaussian 
core.  It is the Gaussian core that we can measure and that is important 
when determining the brightness of a light source or luminosity
in a collision experiment.

The tail-cut is applied by setting the minimum impact parameter as
\begin{equation}
b_{min}=\sqrt{\frac{1}{n\pi\tau_b\nu}},
\end{equation}
where $\tau_b$ is the longest damping time in the bunch frame and $\nu$
is the average particle velocity in the bunch frame.  
If $\epsilon_a$ is greater than $\epsilon_b$ 
and $\frac{\sigma_p\sigma_z}{\gamma^2}$,
then $\nu\approx c\gamma\sqrt{\frac{\epsilon_a}{\beta_a}}$.

The computed IBS growth rate is directly proportional to the Coulomb log 
and is expressed as the logarithm of the maximum impact parameter over the 
minimum,
\begin{equation}
C_{\Lambda}=\log\frac{b_{max}}{b_{min}}.
\end{equation}

In hadron and ion machines, such as the Tevatron and RHIC, the damping time 
is very long and there are enough of even the very 
large-angle scatters to populate
a Gaussian distribution.  A tail-cut does not significantly affect the 
calculated IBS distributions for those machines.
However, for machines with strong damping, such as lepton storage rings, 
very few large-angle scattering events occur per damping time, and applying 
the tail-cut is essential to reliably computing the equilibrium distribution 
of the Gaussian core of the bunch.  In CesrTA, applying the tail-cut 
significantly changes the calculated growth rate.  With the tail-cut, the 
average Coulomb log in CesrTA at $1.6\times10^{10}$ particles/bunch is $9.4$. 
Without the tail-cut, that is, if we assume that the maximum scattering
angle is $90^{\circ}$, the average Coulomb log is $17.6$.

\subsection{Monte Carlo IBS Simulations\label{sec:mc}}
In addition to the analytic IBS calculations discussed above, 
we have developed a  Monte Carlo simulation
based on Takizuka and Abe's plasma collision model \cite{ta:1977}.  
An ensemble of $2000$ particles representing the bunch distribution 
is tracked element-by-element using the {\tt Bmad} standard
tracking methods \cite{bmad:2006}.  
We use an analytic model of the damping wiggler field, which is based
on a fit to a finite element calculation \cite{sagan-wiggler:2003}.
Tracking through wigglers is by symplectic integration.

At each element, the ensemble is converted from canonical to spatial 
coordinates and  boosted into its center of momentum frame 
where the particles are non-relativistic.  Then Takizuka and Abe's
collision model is applied:
\begin{enumerate}
\item The bunch is divided into cells.  This enforces locality.
\item Particles in each cell are paired off.  
Each particle undergoes only one collision.
\item The change in the momentum of the pair 
is calculated, taking into account their 
relative velocities and the density of particles in the cell.
\end{enumerate}
The ensemble is then boosted back to the lab frame 
and transformed back into canonical coordinates.

Note that this is not a Monte Carlo simulation of individual 
scattering events.
Such a simulation would require the calculation of $\frac{N!}{2}$ scattering
events per element and is not computationally feasible.  Takizuka and Abe's 
formalism calculates the expectation value of the change
in the momentum of a test particle traveling through a ``wind'' of nearby 
particles.  The relative velocity of the paired
particles determines the velocity vector of the wind.  The rate of 
change of the particle momentum due to scattering events is 
assumed to be constant through the length of the element.

A log term corresponding to the Coulomb log appears 
in Takizuka and Abe's formalism.  The calculation of the expectation value 
of the change in the momentum of the particles assumes many small-angle 
scattering events.  This method of Monte Carlo simulation is subject to 
the central limit theorem and tail-cut in the same 
way as the analytic calculations.

\subsection{Potential Well Distortion (PWD)}
The bunch interacts with structures in the vacuum system, resulting in 
wake fields that act back on the bunch.  One consequence of this
is a voltage gradient along the length of the bunch.  Particles at the 
head of the bunch lose energy to the vacuum system.  Part of this
energy is reflected back to the tail of the bunch, effectively transferring 
energy from the head of the bunch to the tail.  In machines
that operate above transition, particles with less energy move ahead 
relative to the reference particle, and those with more energy move
back.  The result is bunch lengthening.  The amount of lengthening is 
sensitive to the total bunch charge,
but not to the transverse dimensions of the bunch.

Energy that is reflected back into the bunch does not change the 
total energy of the bunch and is referred to as the inductive ($L$) or
capacitive ($C$) part
of the impedance.  Energy absorbed by the vacuum system does 
change the total energy of the bunch and is referred to as the
resistive part of the impedance ($R$).  

In the general case, the impedance is frequency dependent.
Here, the effect of potential well distortion is approximated as a
current-dependent RF voltage.  The effective RF voltage is \cite{billing:1980}
\begin{equation}
V\left(\tau\right) = V_{rf}\cos\left(\omega\tau+\phi\right) + 
R I_b\left(\tau\right) + L\frac{dI_b\left(\tau\right)}{d\tau},
\end{equation}
where $\tau$ is relative to the bunch center.
The resistive impedance $R$ tends to shift the synchronous phase but 
does not contribute to lengthening.  The inductive part $L$ changes 
the Gaussian profile of the bunch, leading to real bunch lengthening.

In principle, there is also a capacitive part to the impedance.  
Its effect is to shorten the bunch.  In CesrTA, only bunch lengthening is 
observed.  This is because the inductive term in the overall impedance 
is much larger than the capacitive.  Hence, the reactive part of the 
impedance is modeled as entirely inductive.  In theory, the inductive, 
capacitive, and resistive parts of the impedance could each be determined from
the shape of the longitudinal profile of the bunch.
However, our measurements are not detailed enough to determine 
if a capacitive part of the impedance is counteracting the inductive part.

\begin{figure}
\centering
\subfloat{
  \includegraphics[width=0.90\columnwidth]{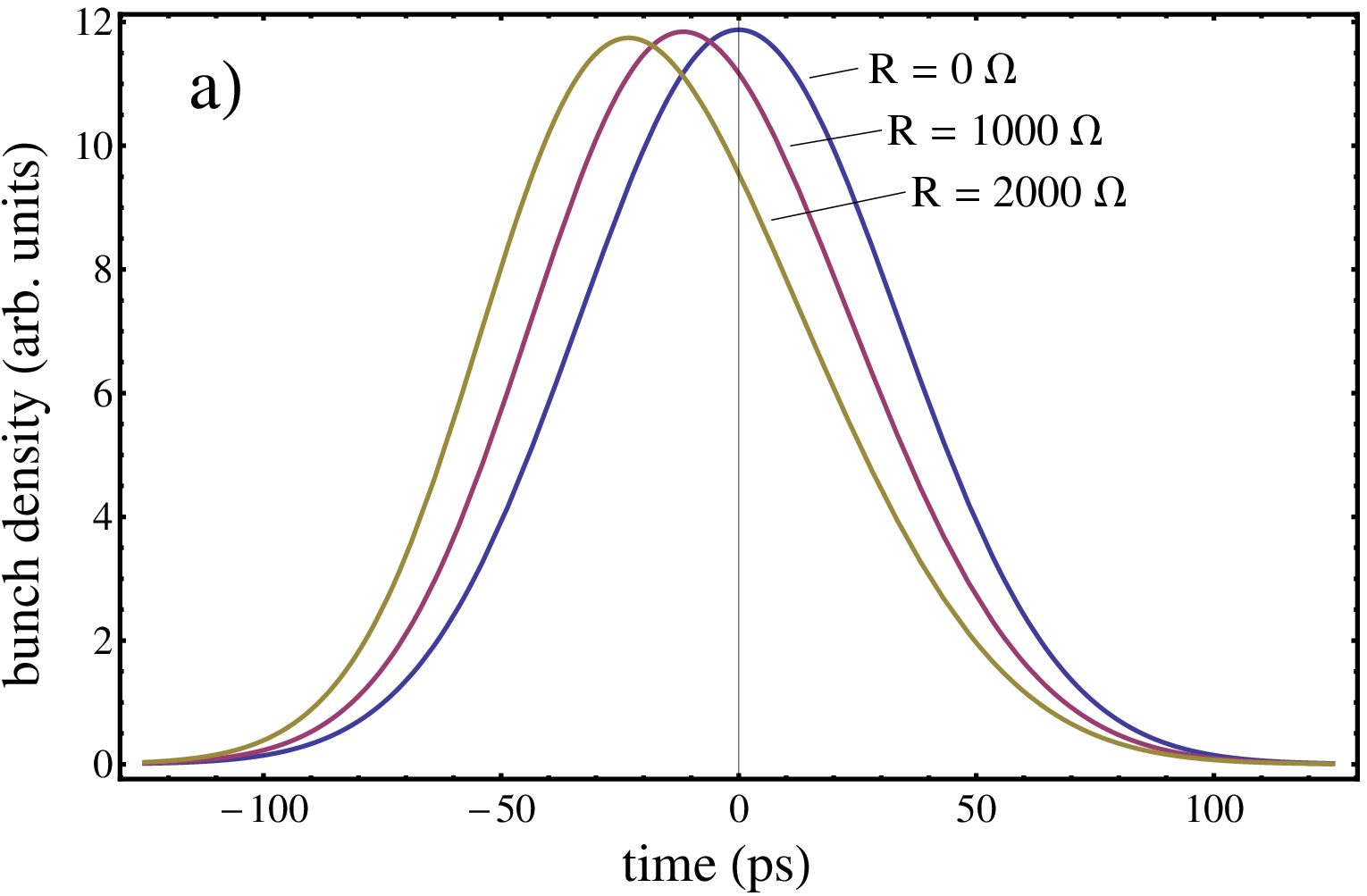}
  \label{fig:rl-pwd:R}
}\\
\subfloat{
  \includegraphics[width=0.90\columnwidth]{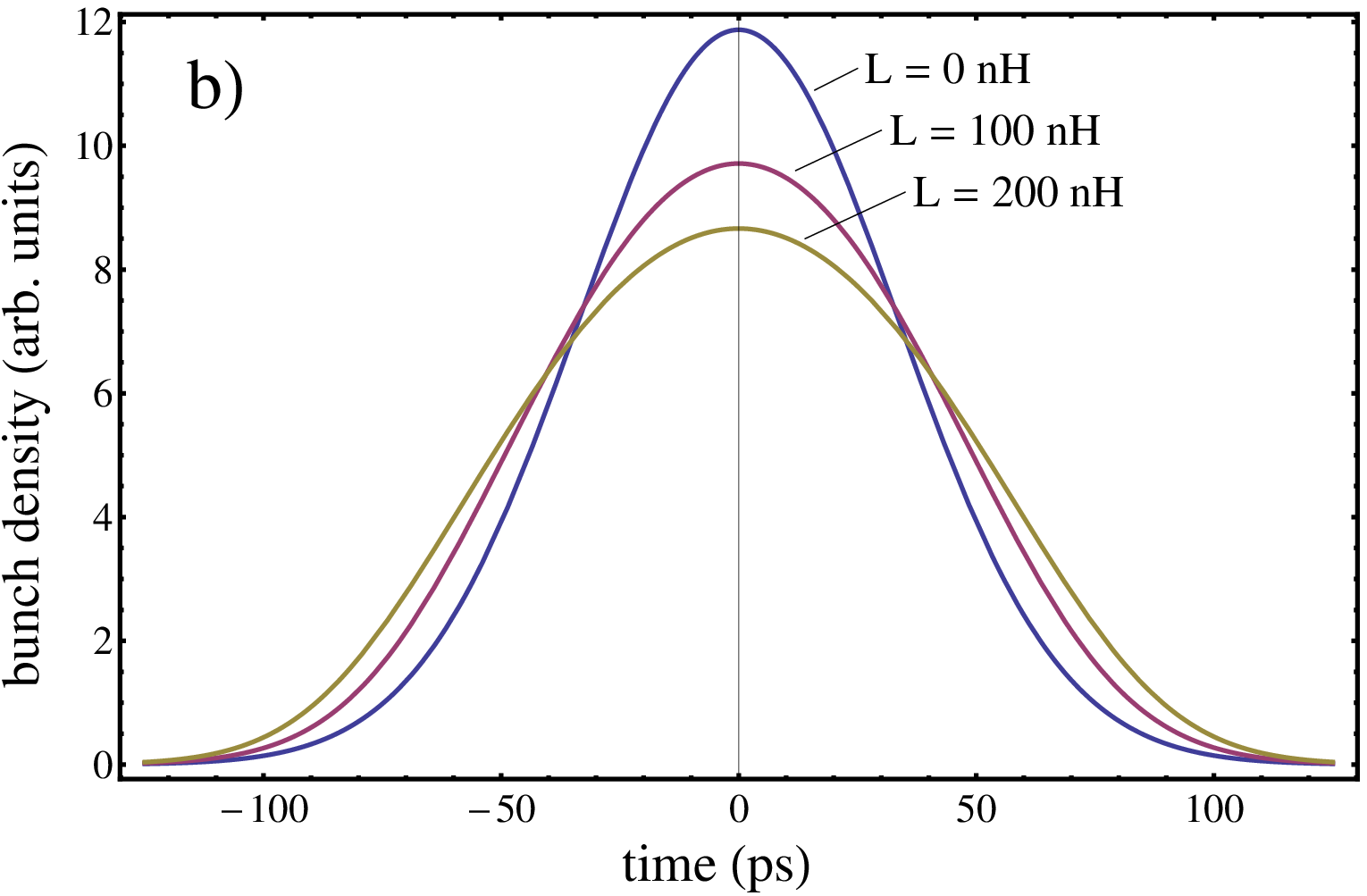}
  \label{fig:rl-pwd:L}
}
\caption{Effect of \protect\subref{fig:rl-pwd:L} inductive and \protect\subref{fig:rl-pwd:R} 
resistive parts of the 
longitudinal impedance on the longitudinal profile of the bunch.
\label{fig:rl-pwd}}
\end{figure}

A derivation of PWD based on Vlassov theory results in a differential 
equation for the longitudinal profile of the bunch \cite{billing:1980},
\begin{equation}
\frac{\partial\psi}{\partial\tau}=-\frac{e E_0 \psi}{\sigma_E^2\alpha T_0}
\left(\frac{ V_{rf}\cos\left(\omega\tau+\phi\right)+Q R\psi-U_0}
{1+\frac{e E_0 Q L \psi}{\sigma_E^2\alpha T_0}}\right),
\label{eqn:pwd-deq}
\end{equation}
where $E_0$ is the beam energy, $\sigma_E$ is energy spread, $\alpha$ is 
momentum compaction, $T_0$ is the period of the ring,
$V_{rf}$ is the total RF cavity voltage, $\omega$ is the RF frequency, 
$\phi$ is the phase of the reference particle with respect
to the RF, $Q$ is the bunch charge, $U_0$ is the energy lost per 
particle per turn, $R$ is the resistive part of the longitudinal
impedance, and $L$ is the inductive part of the longitudinal impedance.  
$\psi\left(\tau\right)$ is the longitudinal profile of the bunch.
Equation (\ref{eqn:pwd-deq}) is used to compute the effect of various 
resistive and inductive impedances on the longitudinal profile of the bunch. 
The results are shown in Fig.~\ref{fig:rl-pwd}.

We have incorporated the effect of PWD in our analytic model of IBS. Equation 
(\ref{eqn:pwd-deq}) is used to compute bunch length, including the energy 
spread resulting from intrabeam scattering.
Comparing the measured bunch length versus current data to the simulation 
result, $L$ is determined to be $25.9^{+18.7}_{-17.2}$~nH for positrons
and $21.1^{+15.3}_{-14.5}$~nH for electrons.  Our bunch 
length predictions are largely insensitive to $R$, and we use 
the published value of $1523$~$\Omega$ given by 
Holtzapple et al. \cite{holtzapple:2000}.  At the time of this 
writing, PWD has not been implemented in the Monte Carlo simulation.
\begin{figure}
\centering
\includegraphics[width=\columnwidth]{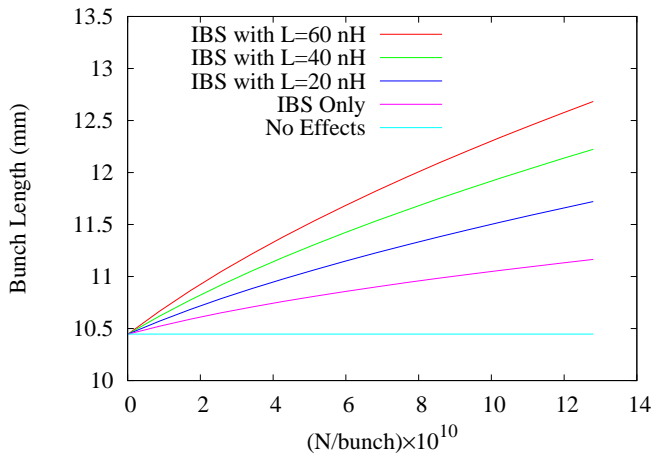}
\caption{Simulated effect on bunch length of PWD in combination with IBS.
\label{fig:pwd-example}}
\end{figure}

As shown in Fig.~\ref{fig:rl-pwd}, resistive impedance has a negligible 
effect on the length of the longitudinal profile, whereas the inductive
impedance $L$ distorts the Gaussian profile and generates bunch 
lengthening.  Figure \ref{fig:pwd-example} shows the contribution of 
the potential well distortion to the bunch length assuming various 
values for the inductive impedance.

The current-dependent energy spread in CesrTA is determined by measuring the 
dependence of the horizontal beam size on the horizontal dispersion at the 
instrument source point.  The dispersion is varied with the help of a closed 
dispersion bump around the source-point.  The 
horizontal beam size is measured under two sets of conditions as the number 
of particles in a single bunch decays
from $1.3\times10^{11}$ down to $2.4\times10^{10}$.
Horizontal dispersion is $2.28$~cm in the first set of 
conditions, and $22.1$~cm in the second.  The measured energy spread
is $\sigma_E/E=\left(8.505\pm0.314\right)\times10^{-4}$ and is independent 
of current within the measurement uncertainty.  The design value of the 
fractional energy spread as determined using the standard radiation integrals 
is $8.129\times10^{-4}$.  There is no evidence of a microwave 
instability, which would appear as an energy spread that increases 
with current above some threshold current.

\subsection{Projected Beam Sizes}
Beam sizes from the simulations are obtained from the $\left<xx\right>$, 
$\left<yy\right>$, and $\left<zz\right>$ elements of the $6 \times 6$ 
beam envelope matrix. These are evaluated at the instrumentation 
source-points.  The beam sizes obtained by this method are the projections 
of the beam into the horizontal, vertical, and longitudinal dimensions and 
are the bunch profiles actually seen by the instrumentation.  This method 
naturally takes into account arbitrary coupling among the 6 normal mode 
phase-space coordinates of the bunch.

\subsection{Method Comparison}
\begin{figure}
\centering
\subfloat{
  \includegraphics[width=0.90\columnwidth]
                  {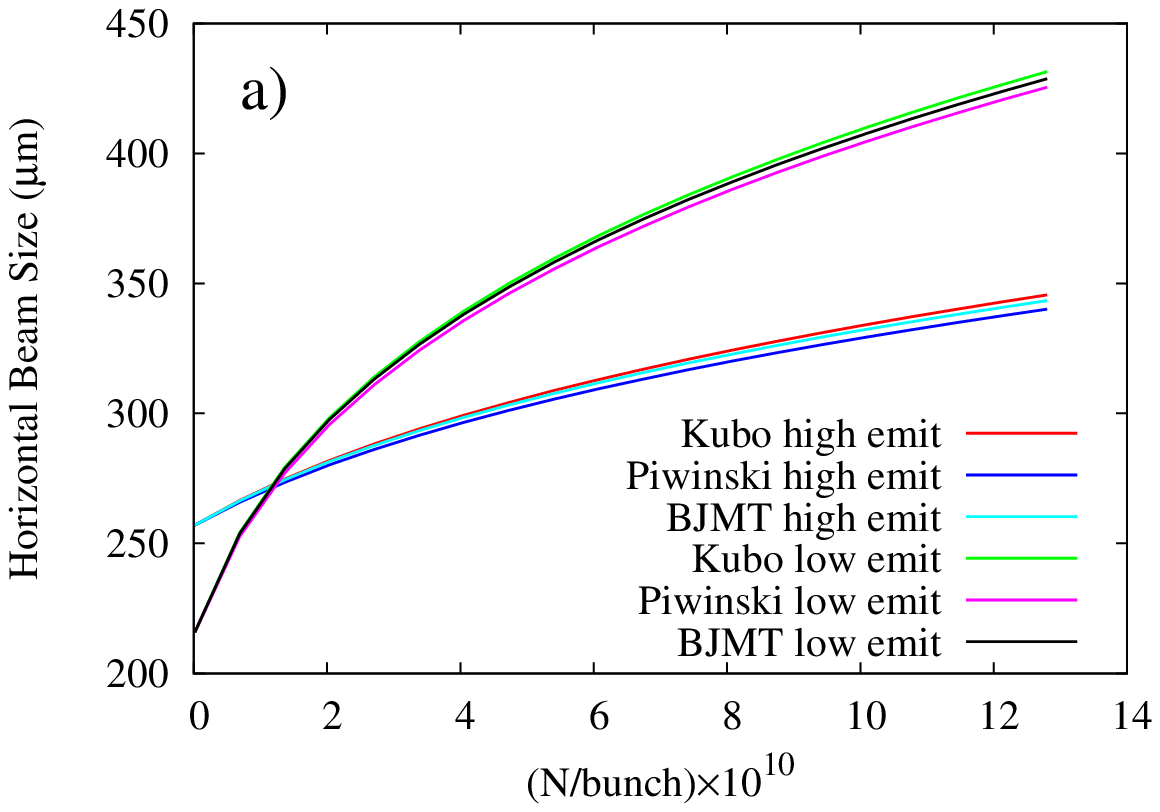}
  \label{fig:Method-Comp-horiz}
}\\
\subfloat{
  \includegraphics[width=0.90\columnwidth]
                  {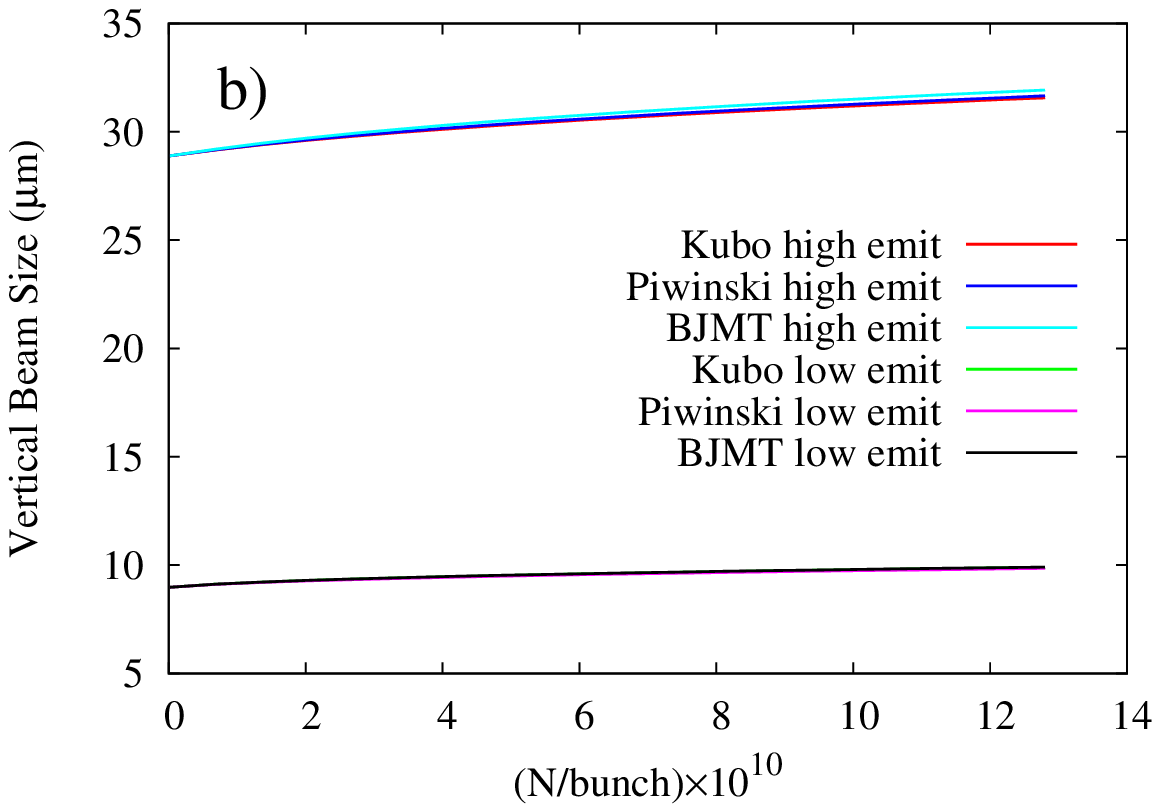}
  \label{fig:Method-Comp-vert}
}\\
\subfloat{
  \includegraphics[width=0.90\columnwidth]
                  {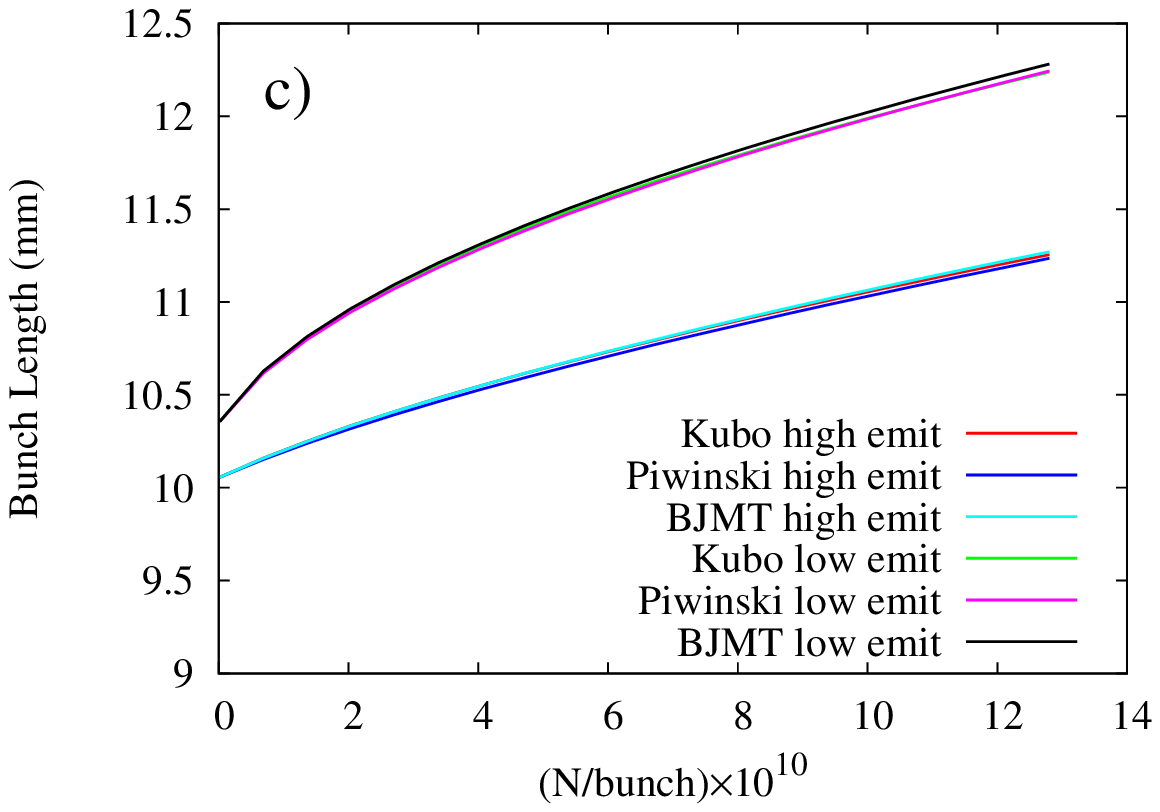}
  \label{fig:Method-Comp-long}
}
\caption{Comparing \protect\subref{fig:Method-Comp-horiz} Horizontal, 
\protect\subref{fig:Method-Comp-vert} vertical,
and \protect\subref{fig:Method-Comp-long} longitudinal beam size versus 
current for Kubo's, Piwinski's, and the Bjorken-Mtingwa (BJMT) IBS
formalisms.  The high emittance lattice 
has $\epsilon_{x0}=4.6$ nm$\cdot$rad,
$\epsilon_{y0}=14.3$~pm$\cdot$rad, and $\sigma_{z0}=10.0$ mm.  
The low emittance lattice has $\epsilon_{x0}=2.8$ nm$\cdot$rad,
$\epsilon_{y0}=1.5$~pm$\cdot$rad, and $\sigma_{z0}=10.3$ mm.
\label{fig:Method-Comp}}
\end{figure}

In addition to Kubo and Oide's method,
two other commonly used methods for calculating IBS growth rates are one
by Bjorken and Mtingwa \cite{bjorken-mtingwa:1983} and a version
of Piwinski's original derivation that includes derivatives of the 
lattice optics \cite{piwinski:1974}.
Figure~\ref{fig:Method-Comp} shows the
equilibrium beam sizes versus current calculated using each of 
the three methods.

We treat the Coulomb log the same way in each method and apply the tail-cut.
Applying the tail-cut to Piwinski's original method requires modifying
the derivation so that the minimum and maximum scattering angles
can be set as parameters.

Bjorken and Mtingwa's and Piwinski's methods are based on
Twiss parameters.
We use normal mode Twiss parameters in place of horizontal, vertical,
and longitudinal
Twiss parameters when evaluating either formalism.
The growth rates given by the formulas are applied to the normal
mode emittances.

These calculations suggest that, provided the Coulomb log is treated
the same, the three most general IBS formalisms predict
similar equilibrium beam sizes.  For the studies shown here, 
transverse coupling is not strong
enough to significantly impact the IBS growth rates.

\section{Current-Dependent Tune Shift}
A current-dependent shift of the coherent tune is observed in CesrTA.  
At $2.1$~GeV, the vertical shift was measured to be $-0.505\pm0.006$~kHz/mA.
The horizontal shift was measured to be $-0.072\pm0.006$~kHz/mA. 
($1$ kHz corresponds to a change in fractional tune of $0.0026$.)
The synchrotron tune has been measured versus current, and no shift 
was observed.  These tune shifts are relevant to IBS studies because the 
beam size will in general depend on proximity of the coherent tune
to resonance lines in the tune plane.  Preparation for IBS studies 
includes identifying a region of the tune plane where the effect of 
resonance lines is minimized for the range of currents to be explored.
The tune plane is scanned with direct measurement as well as tracking 
simulation.  The experimental tune scans are performed by recording the 
beam sizes as the tune is varied by adjusting quadrupole strengths.

\begin{figure}
\centering
\subfloat{
  \includegraphics[width=0.90\columnwidth]{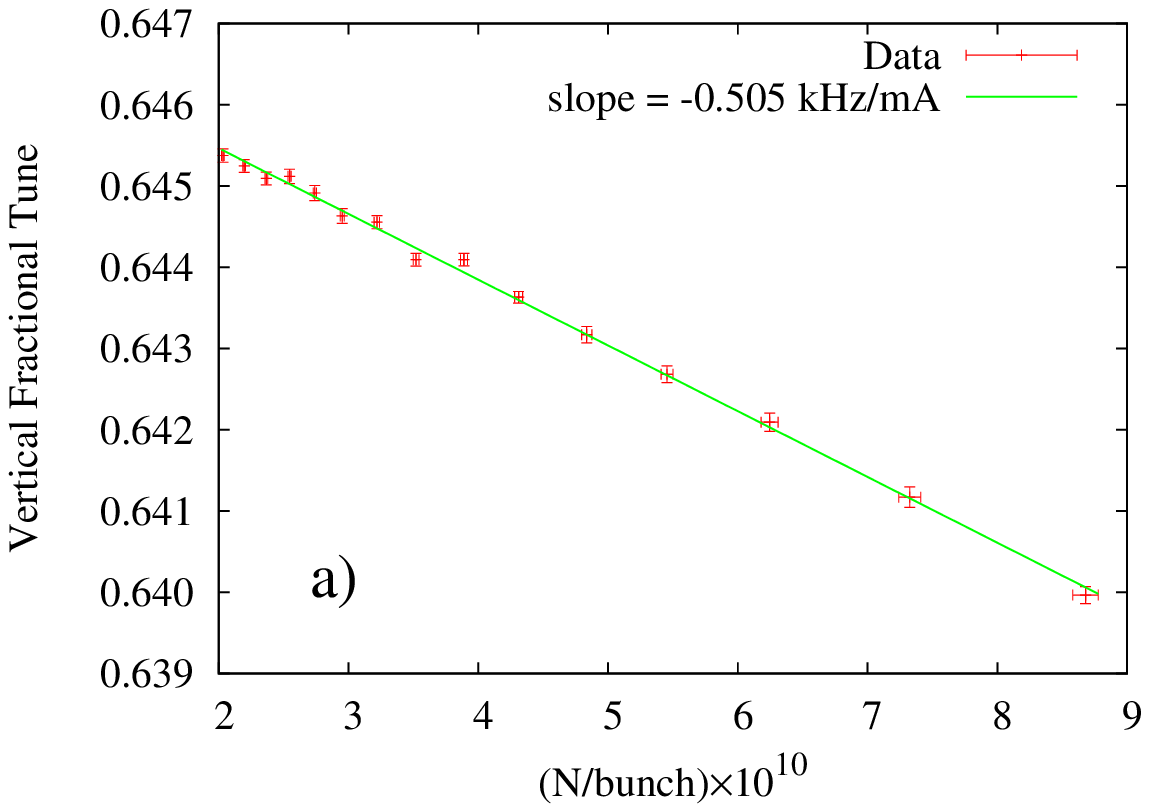}
  \label{fig:68-Tune-Data:v}
}\\
\subfloat{
  \includegraphics[width=0.90\columnwidth]{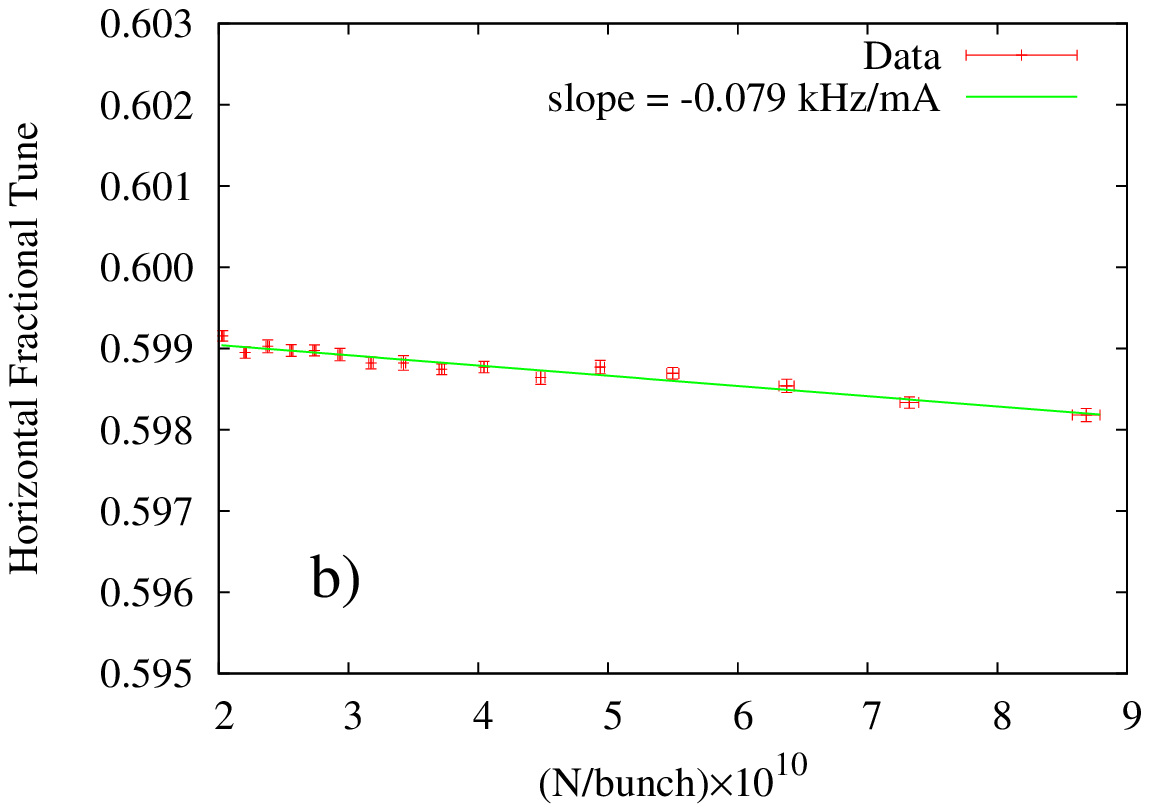}
  \label{fig:68-Tune-Data:h}
}
\caption{Fractional coherent tune versus current for 
\protect\subref{fig:68-Tune-Data:v} the vertical 
and \protect\subref{fig:68-Tune-Data:h} the horizontal plane.  
The resolution of the measurement is $10^{-4}$.  
The revolution frequency is $390.1$ kHz.\label{fig:68-Tune-Data}}
\end{figure}

Figure \ref{fig:68-Tune-Data} shows the measured dependence of vertical 
and horizontal coherent tunes on bunch current.  The betatron frequencies 
are measured via a pair of spectrum analyzers connected to beam position 
monitor (BPM) buttons.  

\begin{figure}
\centering
\includegraphics[width=\columnwidth]{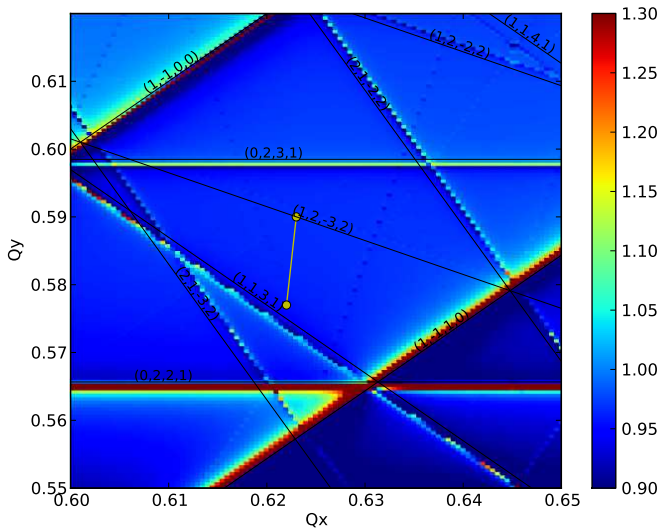}
\caption{Simulated tune scan based on a lattice model that 
includes magnet misalignments and corrector magnet settings 
determined according to our emittance tuning procedure.
The yellow line shows how the coherent tunes increase as a bunch 
decays from $1.3\times10^{11}$ down to to $1.6\times10^{9}$ 
particles.  \label{fig:Sim-Tune-Scan}}
\end{figure}

Figure \ref{fig:Sim-Tune-Scan} shows a simulated tune scan.
The color scale shows the rms value of the vertical-like 
normal mode action $J_b$ of a particle tracked for $2000$ turns, 
normalized by its initial value $J_{b0}$.  The thin lines are analytic 
calculations of the form $r Q_x+s Q_y+t Q_z=n$.  
The labels are of the form $\left(r,s,t,n\right)$.  Amplitude-dependent 
tune-shift causes the resonance lines in the simulation to be offset 
from the analytic calculations.  The initial action $J_{b0}$ of the tracked 
particle is set to be about ten times the equilibrium emittance.
The yellow line shows the range of coherent tune spanned as a bunch
decays from $1.3\times10^{11}$ particles to $1.6\times10^{9}$ particles.  
The upper right hand point is the zero current tune.

The simulated and experimental tune scans are generally only 
in approximate agreement.  The lower order resonances,
such as $\left(1,-1,-1,0\right)$, tend to be much broader in 
the experimental tune scan.  The higher order resonances seen 
in the simulated scan do not appear in the experimental scan.  
The working points for the IBS measurements are chosen with consideration
of both the simulated and measured tune scans; further adjustments
are often needed based on machine behavior.

\section{Simulation Lattices}
An element-by-element description of CesrTA 
is used for the analytic and tracking calculations shown here.  
This description includes quadrupoles, sextupoles, bends, 
steerings, skew quadrupole correctors, wigglers, and
RF cavities.  Systematic multipoles are included for those 
sextupoles which have skew quadrupole or vertical
steering windings.  We use an analytic model of the damping 
wiggler field, which is based on a fit to a finite element 
calculation \cite{crittenden:2005}.  
Tracking through wigglers is by symplectic integration.

The vertical IBS rise time depends on the dispersion. However, vertical 
dispersion is zero for an ideal flat ring.  Vertical dispersion is included
in our analytic IBS calculations by introducing $yz$ coupling into 
the 1-turn transfer matrix.  This is done at each element by augmenting 
the 1-turn transfer matrix before utilizing it in the analytic 
IBS calculation.  The transfer matrix $\mathbf{T}$ is replaced with 
with $\mathbf{\tilde{T}}$, where $\mathbf{\tilde{T}}=\mathbf{T}\mathbf{W}$, 
and
\begin{equation}
\mathbf{W}=
\begin{pmatrix}
1& 0& 0& 0& 0& 0\\
0& 1& 0& 0& 0& 0\\
0& 0& 1& 0& 0& -\tilde{\eta}_y\\
0& 0& 0& 1& 0& -\tilde{\eta}'_y\\
0& 0& \tilde{\eta}_y'& -\tilde{\eta}_y& 1& 0\\
0& 0& 0& 0& 0& 1
\end{pmatrix}.
\end{equation}
This transformation preserves the symplecticity of the transfer matrix.  
$\tilde{\eta}_y$ and $\tilde{\eta}'_y$ are dispersion-like
quantities.  An ideal lattice modified according to the above prescription 
with $\tilde{\eta}_y=0.01$ m and $\tilde{\eta}'_y=0.002$ has an 
rms vertical dispersion of $10.9$~mm and a vertical IBS risetime 
similar to that of a lattice with an rms vertical dispersion of $10$~mm.

The vertical dispersion in CesrTA is measured to be less than $15$~mm.  
The upper bound is limited by the resolution of our measurement technique.  
The coupling is determined by direct measurement to be 
$\bar{C}_{12}<0.003$, using an extended Edwards-Teng 
formalism \cite{sagan-rubin:1999}.  This amount of coupling is negligible, 
and the simulation assumes no transverse coupling.

The analytic simulation takes the measured low current
horizontal and vertical beam sizes and bunch length
as input parameters and computes the current dependence.
The horizontal 
emittance used in the calculation is chosen to match the measured near zero 
current emittance.  The vertical emittance is also set to agree with the 
measurement extrapolated to zero current.  (The vertical emittance of 
the design simulation lattice is zero.)  The energy spread and bunch length 
used in the simulation are obtained by evaluating 
the standard radiation integrals.  

The Monte Carlo simulation includes photon emission and so 
requires a realistic vertical dispersion function.  This is generated
by applying a distribution of misalignments to the ideal lattice, then 
correcting the phase advance, coupling, orbit, and vertical dispersion 
according to the same procedure that is applied to CesrTA low-emittance 
tuning \cite{shanks:2011}.  The magnitude of the misalignments is set such 
that the zero current vertical emittance is roughly $15$ pm-rad.

\section{Experiment}
\begin{figure}[h]
\begin{center}
\includegraphics[width=1.00\columnwidth]{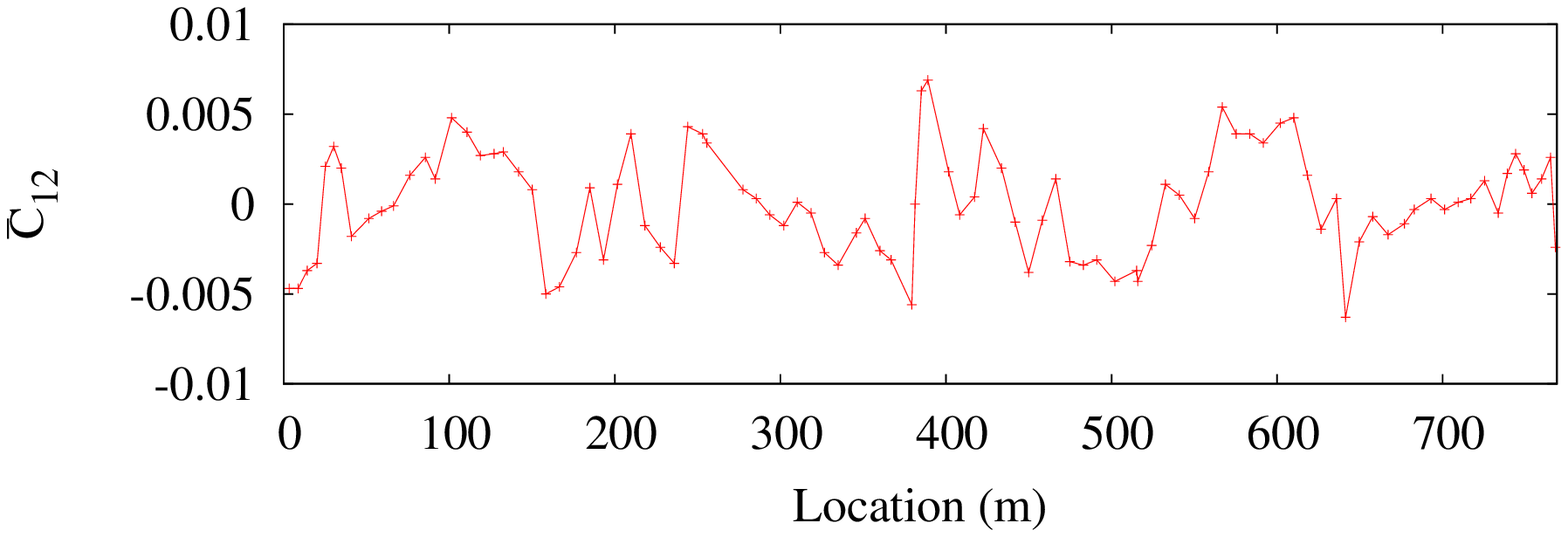}
\includegraphics[width=1.00\columnwidth]{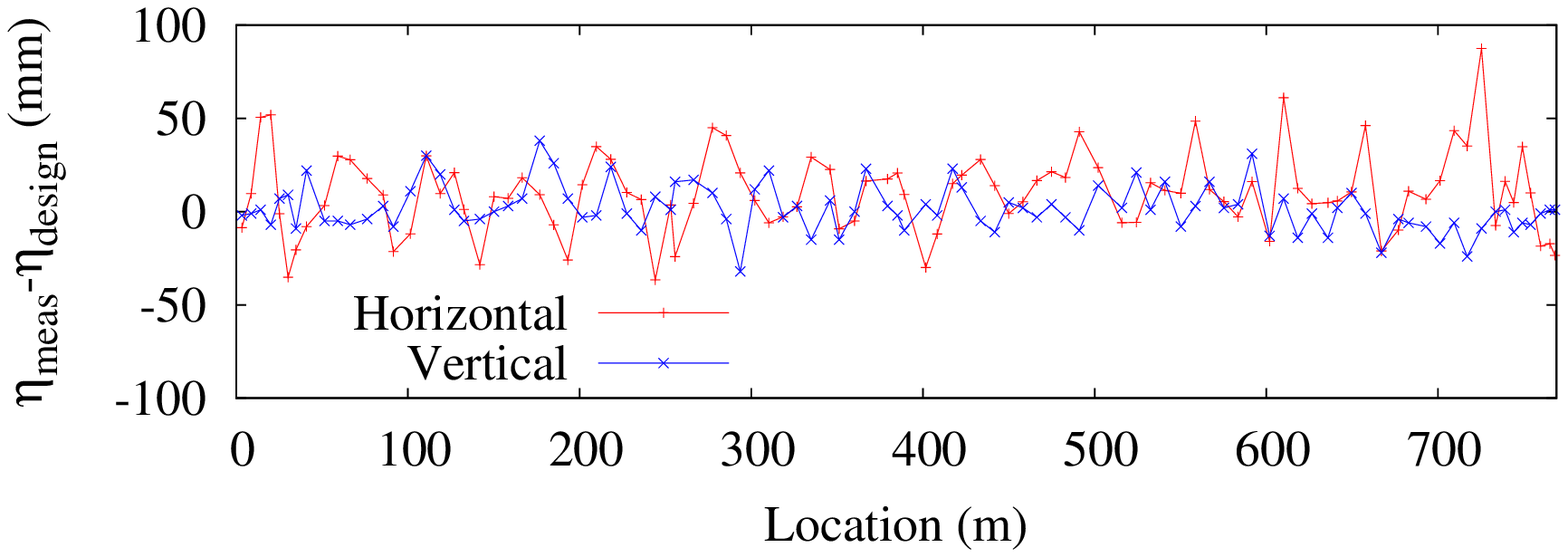}
\includegraphics[width=1.00\columnwidth]{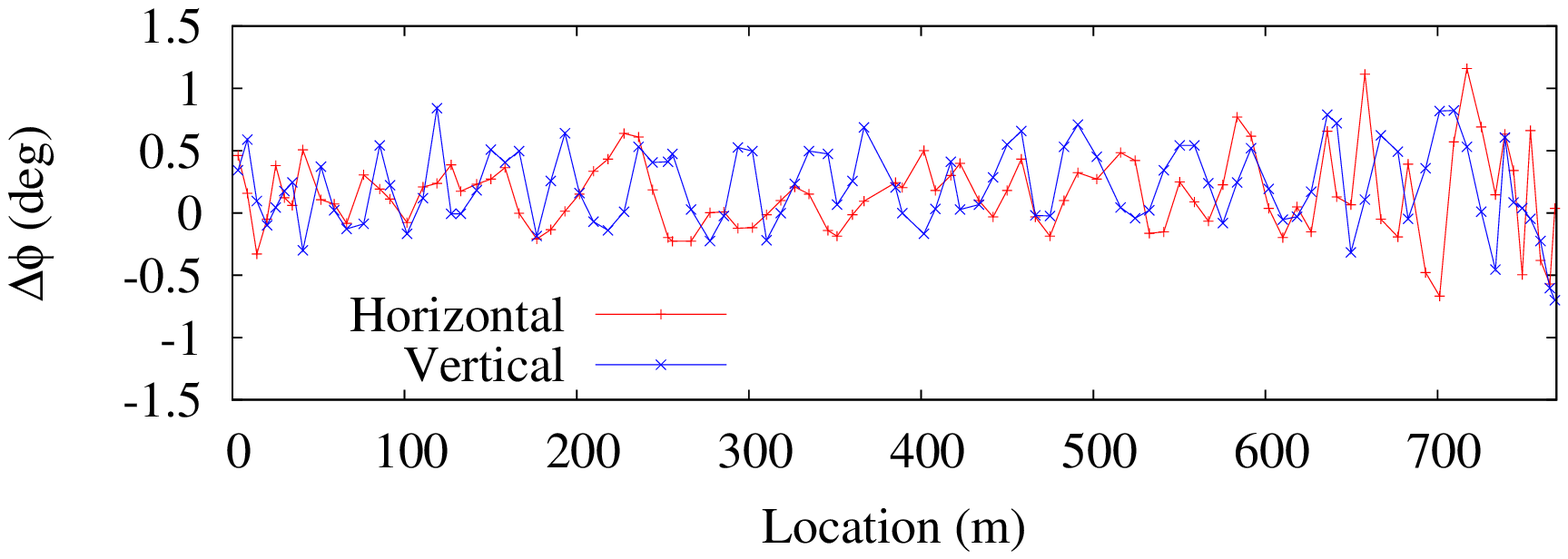}
\caption{Top: Measured transverse coupling at each beam position
monitor after correction. $\bar{C}_{12}$ is the ratio of the
normalized amplitude of vertical to
horizontal motion when the beam is driven at the $a$-mode (horizontal)
tune. 
Middle: Measured dispersion error with respect to design 
$\left(\eta_{measured}-\eta_{design}\right)$.
Bottom: The residual betatron 
phase error $\left(\phi^{h,v}_{measured}-\phi^{h,v}_{design}\right)$.
A $1^\circ$ rms phase error corresponds to a $\sim2$\% $\beta$ error. 
\label{fig:example_corrected}}
\end{center}
\end{figure}

For measurements of intrabeam scattering, we load a specific lattice 
configuration into the storage ring, which includes beam energy, working 
point, and RF voltage.  The orbit, betatron phase,  transverse coupling, 
and dispersion are measured at each of the $100$ beam position monitors 
around the ring and then corrected to match the design.  The phase and 
coupling data is derived from turn by turn position measurements at each 
of the beam position monitors for a resonantly excited beam \cite{saganpc}.
The measurement of betatron phase and coupling takes $10$ seconds,
with phase reproducibility of order $0.05$ deg and $\bar {C}_{12}$
reproducibility of $0.002$.  The dispersion is determined by directly measuring 
the dependence of closed orbit on beam energy (RF frequency).  The machine 
model is fit to the measured phase, coupling and dispersion by varying all 
$100$ quadrupoles, $25$ skew quadrupole, and $55$ vertical steering 
correctors.  The machine optics are forced to match the design by loading 
the fitted magnet changes with opposite sign.  The procedure typically 
converges in a few iterations.  An example measurement of the optical 
functions after correction and just prior to an IBS run is shown in 
Fig. \ref{fig:example_corrected}.  The machine is tuned for minimum 
vertical emittance according to the algorithm given in \cite{shanks:2011}. 
For experiments requiring a larger beam size, the vertical emittance is 
increased by adjusting a closed coupling and vertical dispersion bump that
propagates vertical dispersion through the wigglers.

A single bunch of about $1.6\times 10^{11}$ particles ($10$~mA)
is allowed to decay.
The measurements include
horizontal and vertical beam sizes, streak camera measurements of the 
longitudinal profile, and tunes in all three dimensions.
The beam current decays from $10$~mA to $1$~mA in about $20$~minutes.
The short beam lifetime is due to Touschek scattering; below $1$~mA,
where the charge density is considerably smaller, the beam lifetime
improves significantly.  In the interest 
of time, a large-amplitude pulsed orbit bump is used to scrape particles 
out of the beam in $0.25$~mA decrements.  The discontinuities in the 
data at bunch charge $< 2\times10^{10}$ particles correspond to the 
regime where beam is scraped out.

IBS measurements are taken during dedicated periods of CesrTA
operation.
The IBS measurements in 2011 led us through iterative
improvements in our understanding of how to operate the accelerator
and how to measure IBS effects.  Improvements on the accelerator side
included a better understanding of the tunes and the selection of the
working point (tunes as determined by lattice optics), a better
understanding of the coupling and its impact on the measurements, and
the development of more exact procedures for establishing the desired
machine configuration.  Improvements to the instrumentation included
the implementation of beam size measurements for both electrons and
positrons and the development of more accurate and robust analysis
software.  The data presented here were taken during the April 2012
CesrTA run.  The measurements in December 2012 and April 2013
corroborate and expand on the April 2012 run and have been published
in \cite{ipac13:ibs} and \cite{ipac13:v15}.

\begin{figure}
\centering
\subfloat{
  \includegraphics[width=0.90\columnwidth]{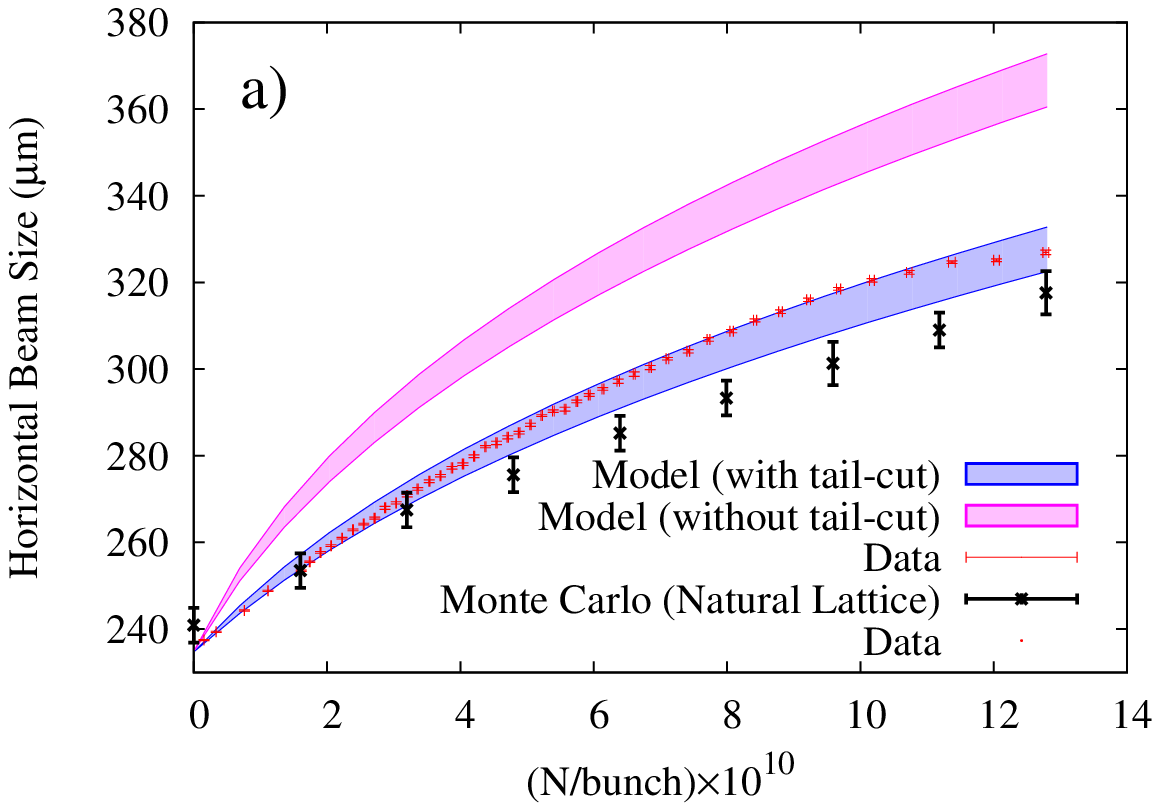}
  \label{fig:4B-Methods-horiz}
}\\
\subfloat{
  \includegraphics[width=0.90\columnwidth]{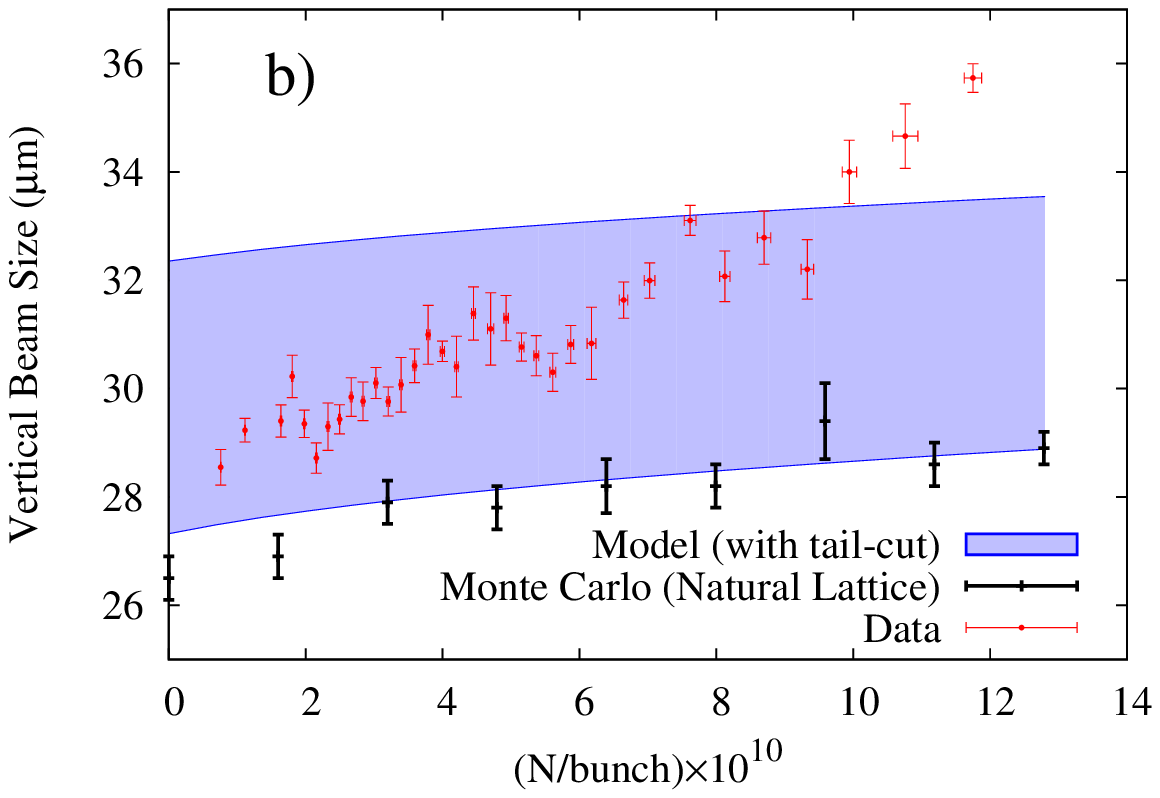}
  \label{fig:4B-Methods-vert}
}\\
\subfloat{
  \includegraphics[width=0.90\columnwidth]{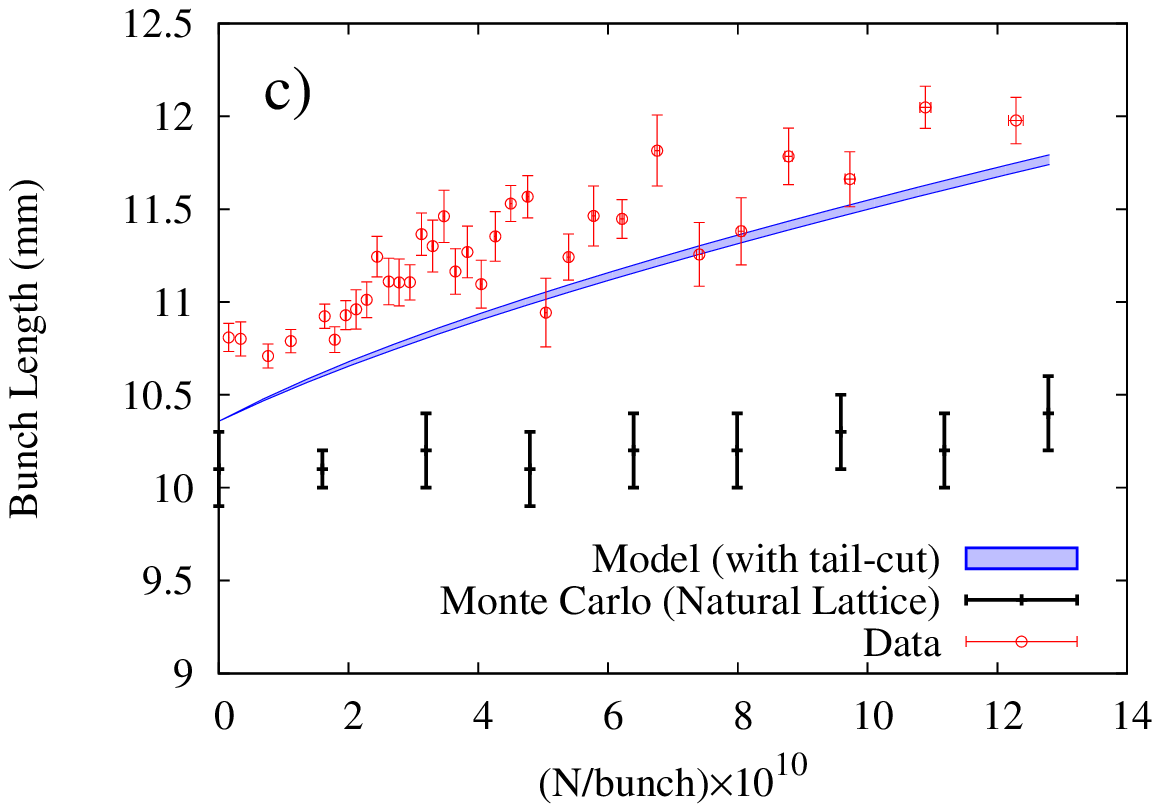}
  \label{fig:4B-Methods-long}
}
\caption{\protect\subref{fig:4B-Methods-horiz} Horizontal, 
\protect\subref{fig:4B-Methods-vert} vertical,
and \protect\subref{fig:4B-Methods-long} longitudinal beam size versus 
current for $e^+$ bunch in conditions tuned for 
minimum vertical emittance.
\label{fig:4B-Methods}}
\end{figure}

Figure \ref{fig:4B-Methods} shows the data from a positron bunch in 
conditions tuned for minimum vertical emittance.
Analytic results from the $\mathbf{\Sigma}$-matrix formalism described in 
Sec. \ref{sec:theory} and the Monte Carlo simulation described
in Sec. \ref{sec:mc} are shown along with the data.  
Some of the error bars in Fig.~\ref{fig:4B-Methods} are below
the resolution of the plot.
The approximate statistical uncertainties at high current are 
shown in Table~\ref{tab:stat-unc}.
\begin{table}[tb]
\centering
\caption{Approximate statistical uncertainties at high current.
\label{tab:stat-unc}}
\begin{tabular*}{\columnwidth}{@{\extracolsep{\fill}}lr}
\hline
\hline
Measurement                       & Uncertainty\\
\hline
Current (horiz. binning)         & $0.3$\%\\
Current (bunch length binning)   & $0.9$\%\\
Horizontal Size                  & $0.2$\%\\
Bunch Length                     & $1.0$\%\\
Vertical Size                    & $0.2$\%\\
\hline
\hline
\end{tabular*}
\end{table}

The systematic uncertainty in the measured horizontal beam size is
about $2$\%, and is due to vibration of optical elements and horizontal
beam motion.  The systematic uncertainty in the vertical measurement is
about $\pm 2$~$\mu$m and is dominated by our understanding of the x-ray 
optics and detector.

The accuracy of the simulation is limited by the ambiguity of
the Coulomb log 
and limited knowledge of the zero current vertical beam size of
the machine.  The simulation result shown here follows the usual 
convention for the tail-cut of $1$ event/damping time as the
cutoff.

The $\mathbf{\Sigma}$-matrix IBS simulation is 
run twice, once with a zero current vertical emittance that extends to
the bottom range of the measurement systematic uncertainty, 
and once that
extends to the upper range of the measurement systematic uncertainty.
The shaded region is the area between those two results.  
This serves two purposes.  First, it reflects the systematic uncertainty
in the vertical beam size measurements.
Second, it gives the reader an idea of how the horizontal
simulation result depends upon particle density as determined by
the vertical beam size.

The zero current vertical emittances
that bound the data in Fig. \ref{fig:4B-Methods-vert}
are $17.9$~pm and $25.1$~pm.  
The shaded regions of \ref{fig:4B-Methods-horiz}
and \ref{fig:4B-Methods-long} show how the horizontal
and vertical simulation results
change as the zero current vertical emittance is varied from the
lower bound to the upper bound.

The measured zero current horizontal emittance, which is an input
parameter to the simulation,
is $3.8$~nm-rad.  For the bunch length and energy spread,
we use the values calculated from the radiation integrals.

The simulation uses a perfectly aligned CesrTA lattice.  
Vertical dispersion is included by modifying the 1-turn transfer
matrix with $\mathbf{W}$ before passing it to the IBS rise-time
calculation.  $\tilde{\eta}$ is set to $10$~mm.

\begin{figure}
\includegraphics[width=\columnwidth]{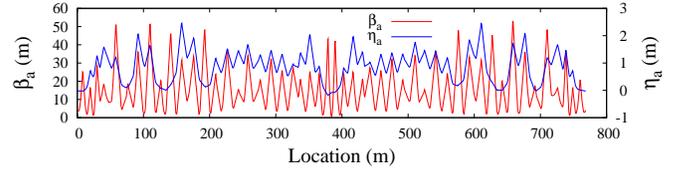}
\caption{CesrTA design $a$-mode (horizontal-like) $\beta$ and 
dispersion $\eta$.
\label{fig:twiss-h}}
\end{figure}
The horizontal emittance
increases from $3.8$ nm-rad at 
low current ($< 1.5\times10^9$ particles/bunch) 
to $10.4$ nm-rad at $1.3\times10^{11}$ particles/bunch.
The reason for the relatively large horizontal blow-up is the large 
horizontal dispersion in CesrTA.  The lattice functions $\beta_a$ 
and $\eta_a$ are shown in Fig.~\ref{fig:twiss-h}.
The rms horizontal dispersion, $\eta_a$, is $1.0$ m and peaks at $2.5$ m.
For comparison, the rms vertical dispersion is less than $15$~mm.
\begin{figure}
\centering
\subfloat{
  \includegraphics[width=0.90\columnwidth]{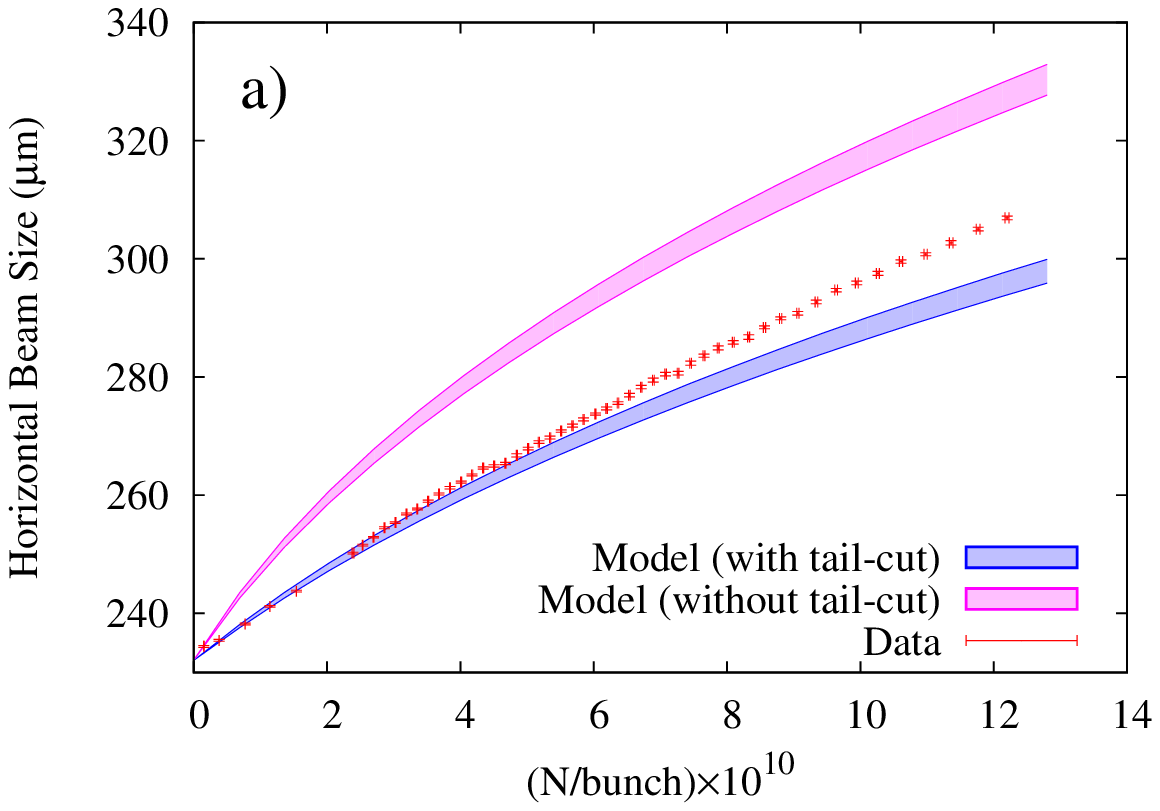}
  \label{fig:53-Data-horiz}
}\\
\subfloat{
  \includegraphics[width=0.90\columnwidth]{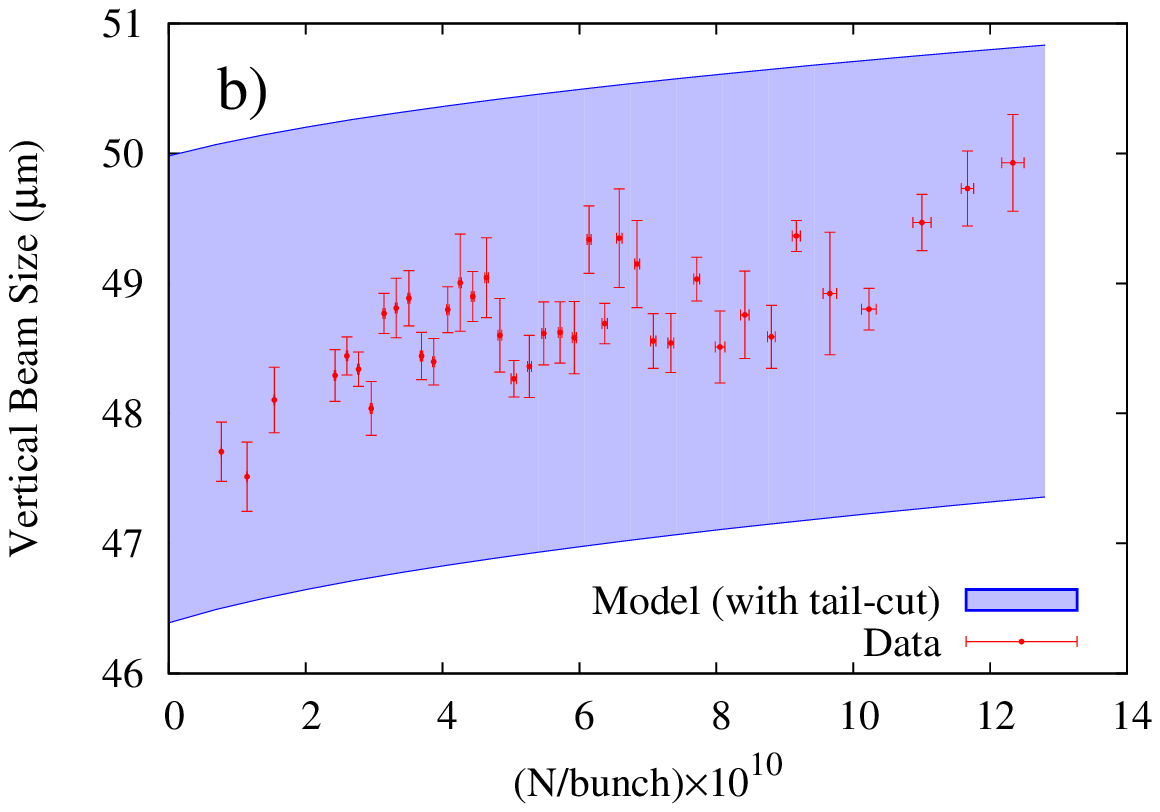}
  \label{fig:53-Data-vert}
}\\
\subfloat{
  \includegraphics[width=0.90\columnwidth]{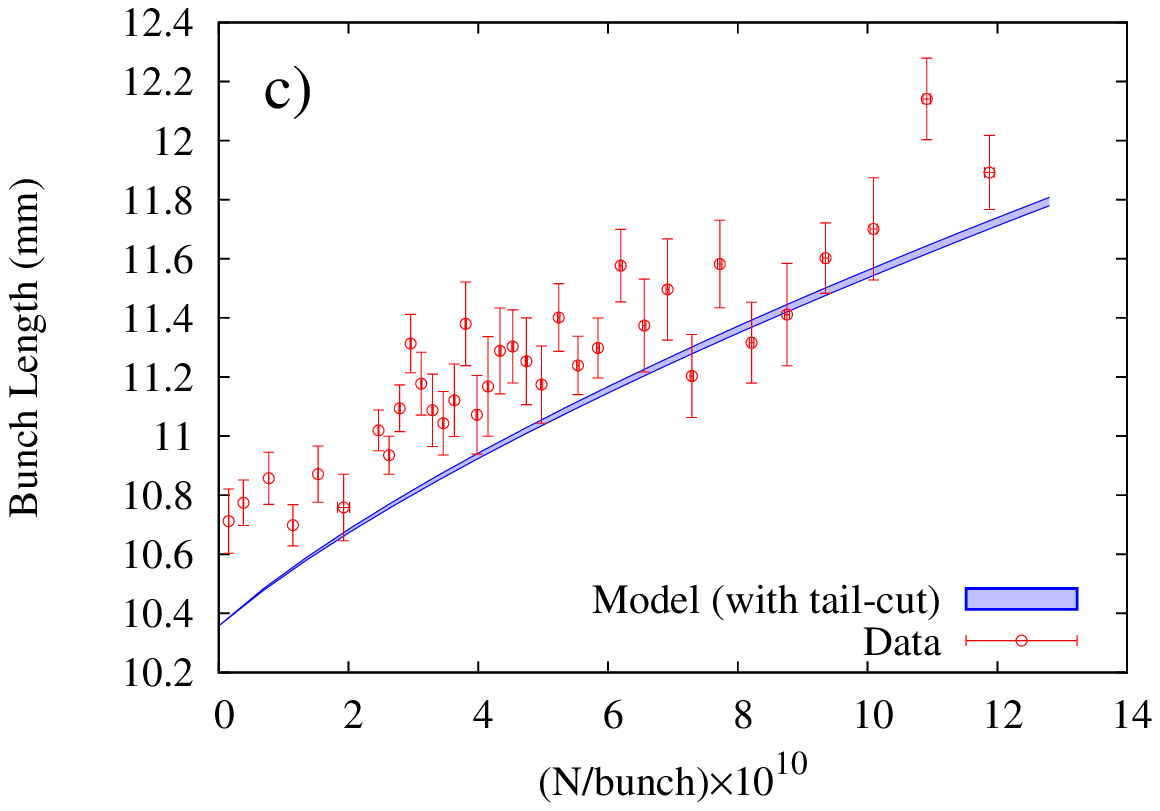}
  \label{fig:53-Data-long}
}
\caption{\protect\subref{fig:53-Data-horiz} Horizontal, 
\protect\subref{fig:53-Data-vert} vertical,
and \protect\subref{fig:53-Data-long} longitudinal 
beam size versus current for $e^+$ bunch with increased
zero current vertical emittance.
\label{fig:53-Data}}
\end{figure}

In Fig.~\ref{fig:53-Data} the zero current vertical emittance of the bunch 
was increased by propagating vertical dispersion through the damping 
wigglers with the help of a closed coupling and dispersion bump.
The larger vertical beam size decreases the particle density, 
which in turn reduces the amount by which IBS blows up the 
horizontal beam size.
The zero current horizontal emittance is $3.7$ nm-rad.
The zero current vertical emittances that bound the data
are $51.6$~pm and $59.9$~pm.

\begin{figure}
\centering
\subfloat{
  \includegraphics[width=0.90\columnwidth]{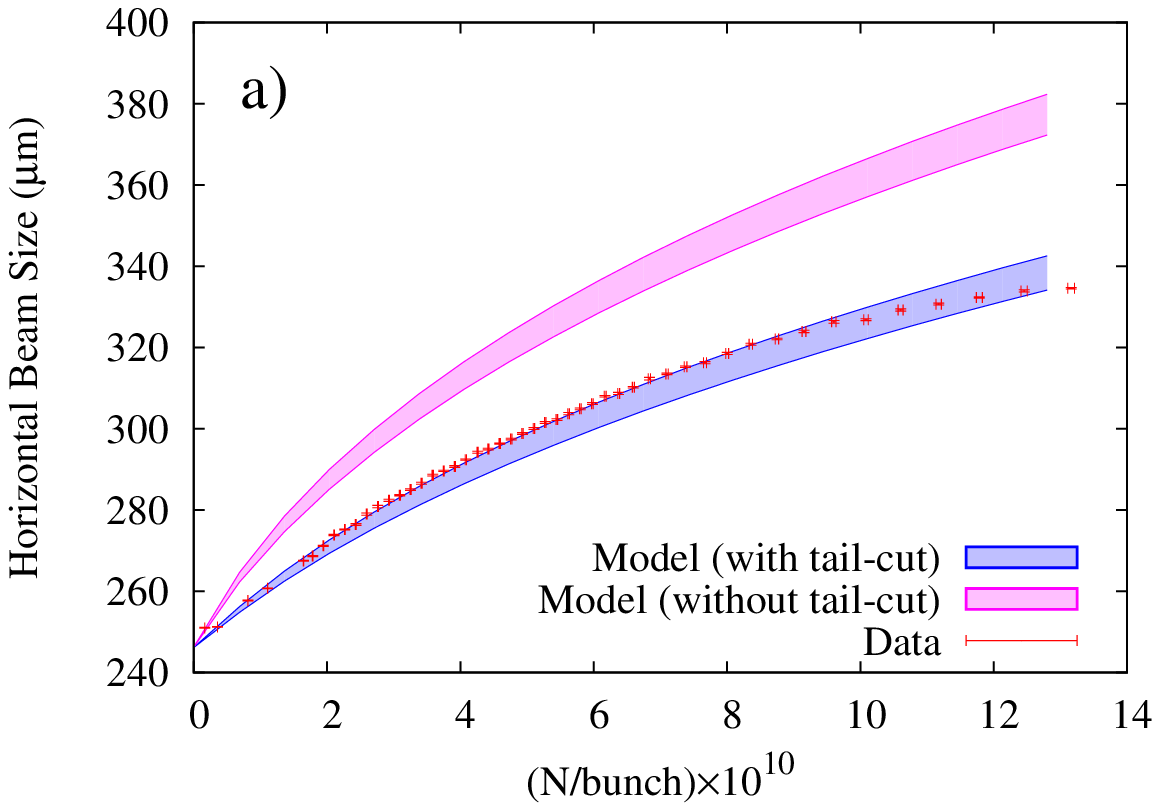}
  \label{fig:68-Data-horiz}
}\\
\subfloat{
  \includegraphics[width=0.90\columnwidth]{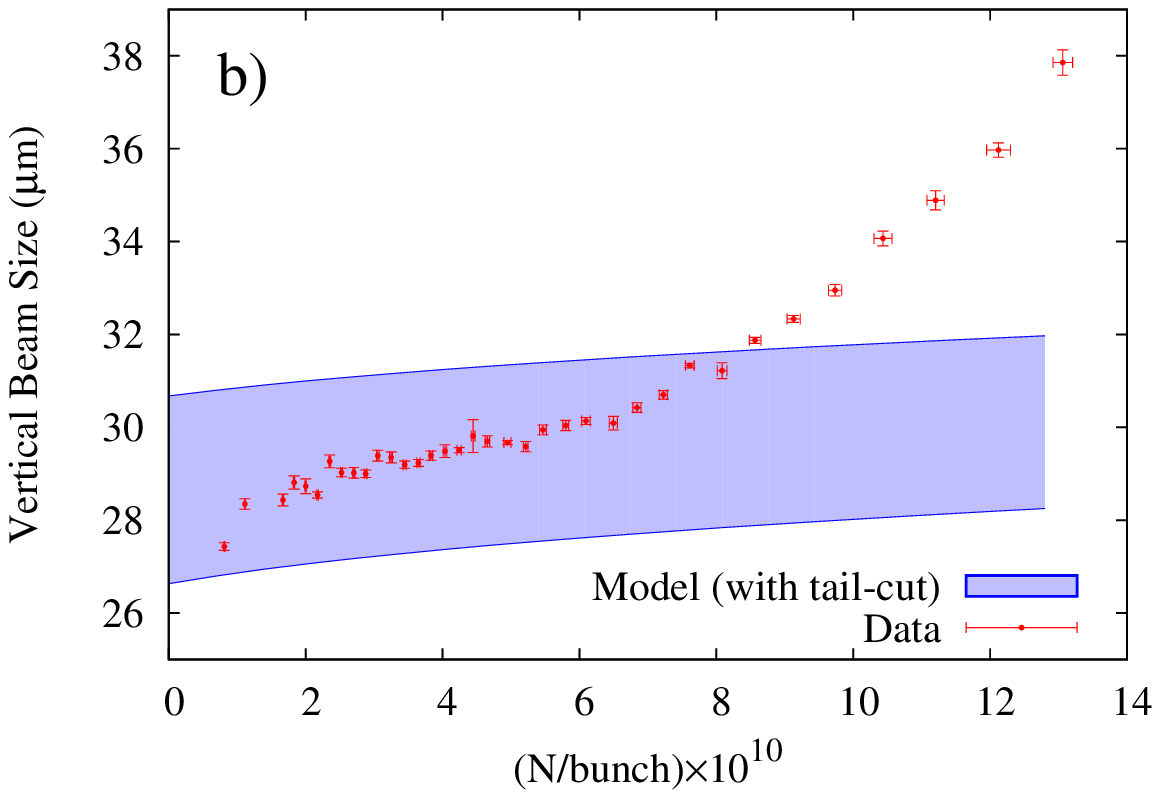}
  \label{fig:68-Data-vert}
}\\
\subfloat{
  \includegraphics[width=0.90\columnwidth]{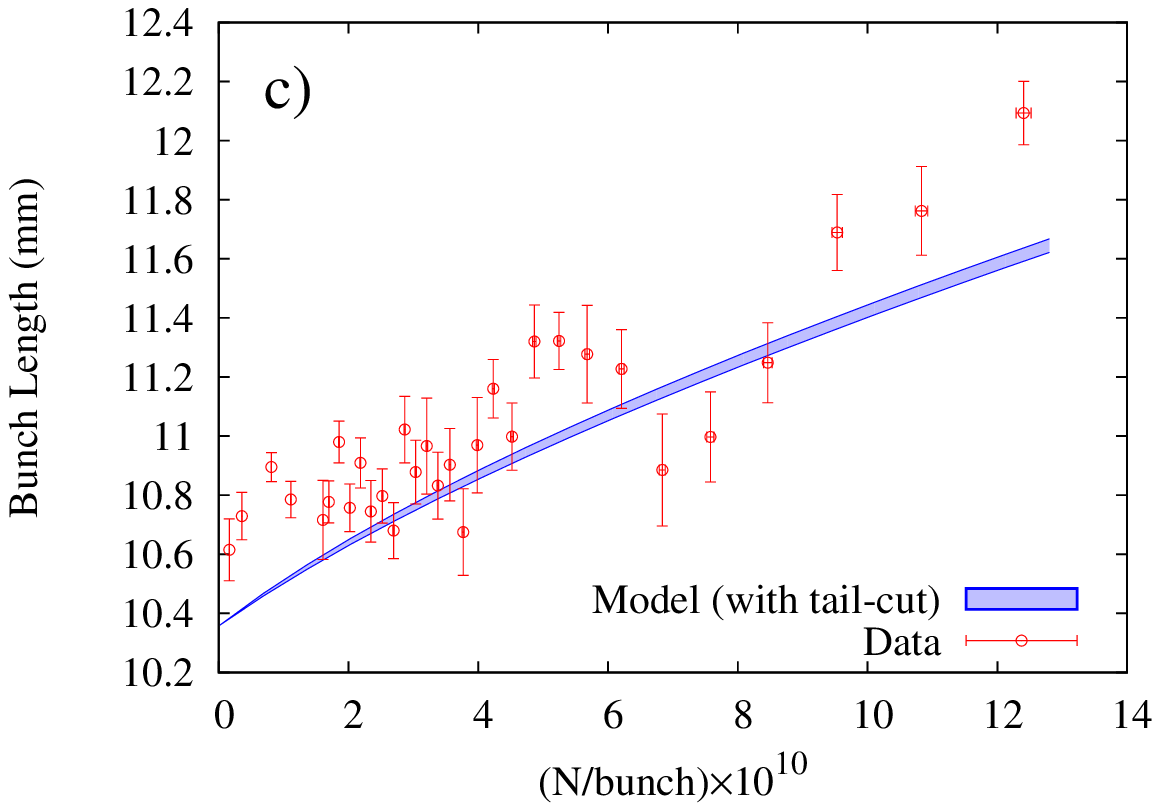}
  \label{fig:68-Data-long}
}
\caption{\protect\subref{fig:68-Data-horiz} Horizontal, \protect\subref{fig:68-Data-vert} 
vertical, and \protect\subref{fig:68-Data-long} longitudinal 
beam size versus current for $e^-$ bunch in conditions 
tuned for minimum vertical emittance. 
\label{fig:68-Data}}
\end{figure}

IBS theory is species-independent.  Measurements of both $e^-$ and $e^+$ 
can help identify machine and instrumentation systematics and distinguish 
IBS from species-dependent beam physics such as electron cloud and ion 
effects.
Figure \ref{fig:68-Data} shows data from an 
electron bunch in conditions tuned for minimum vertical emittance.
The measured horizontal emittance is $4.3$ nm-rad at 
zero current and $8.2$ nm-rad at $4.8\times10^{10}$ particles/bunch.  
The zero current vertical emittances that bound 
the data are $16.9$~pm and $22.4$~pm.

\begin{figure}
\centering
\subfloat{
  \includegraphics[width=0.90\columnwidth]{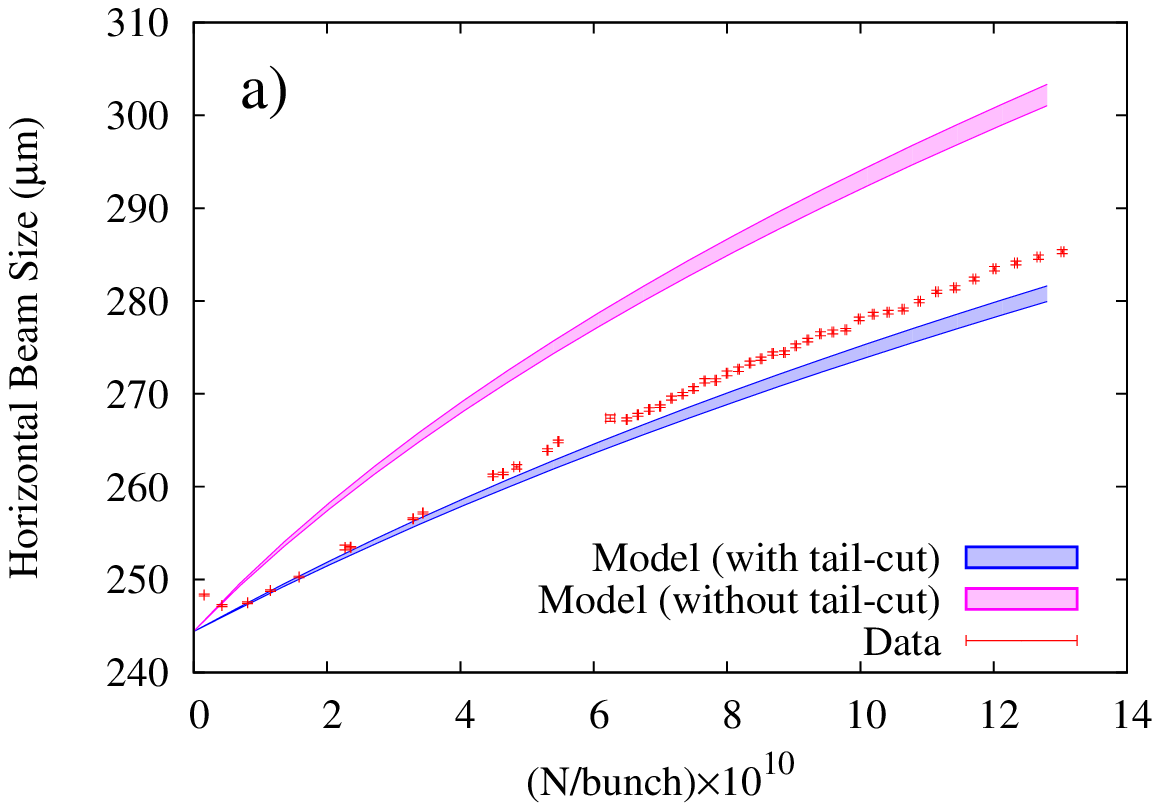}
  \label{fig:6B-Data-horiz}
}\\
\subfloat{
  \includegraphics[width=0.90\columnwidth]{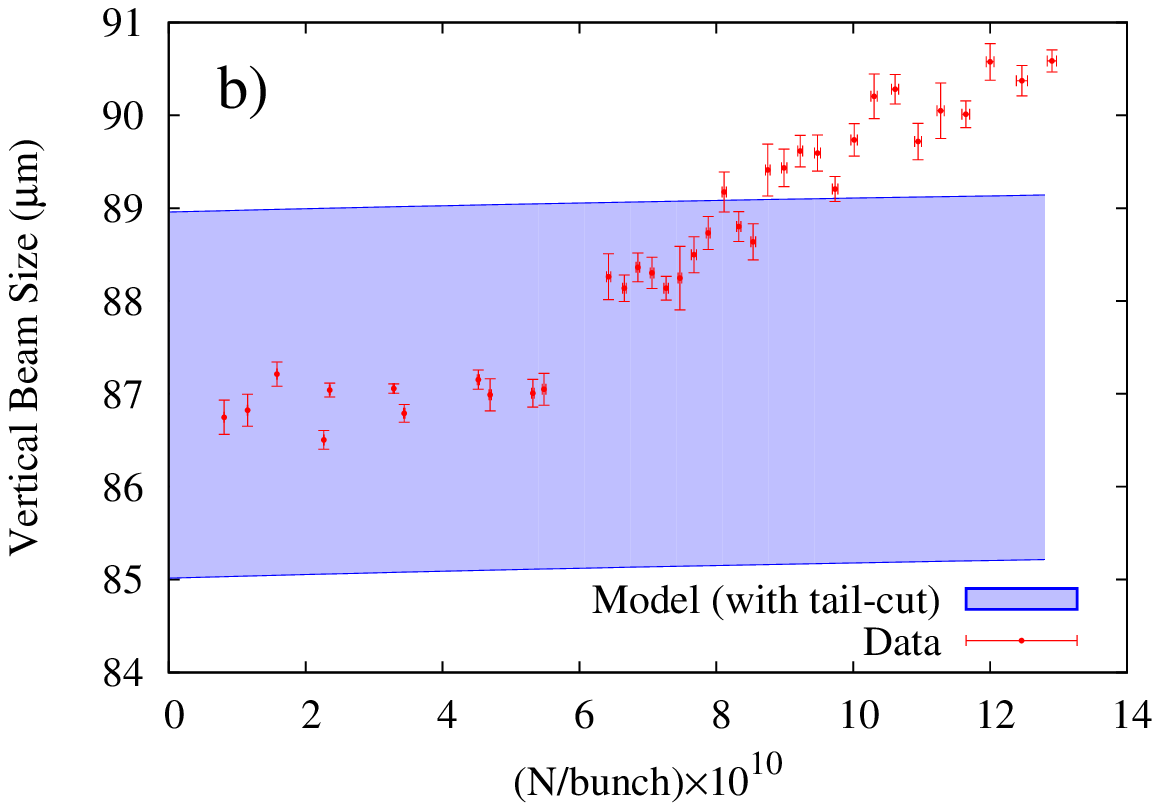}
  \label{fig:6B-Data-vert}
}\\
\subfloat{
  \includegraphics[width=0.90\columnwidth]{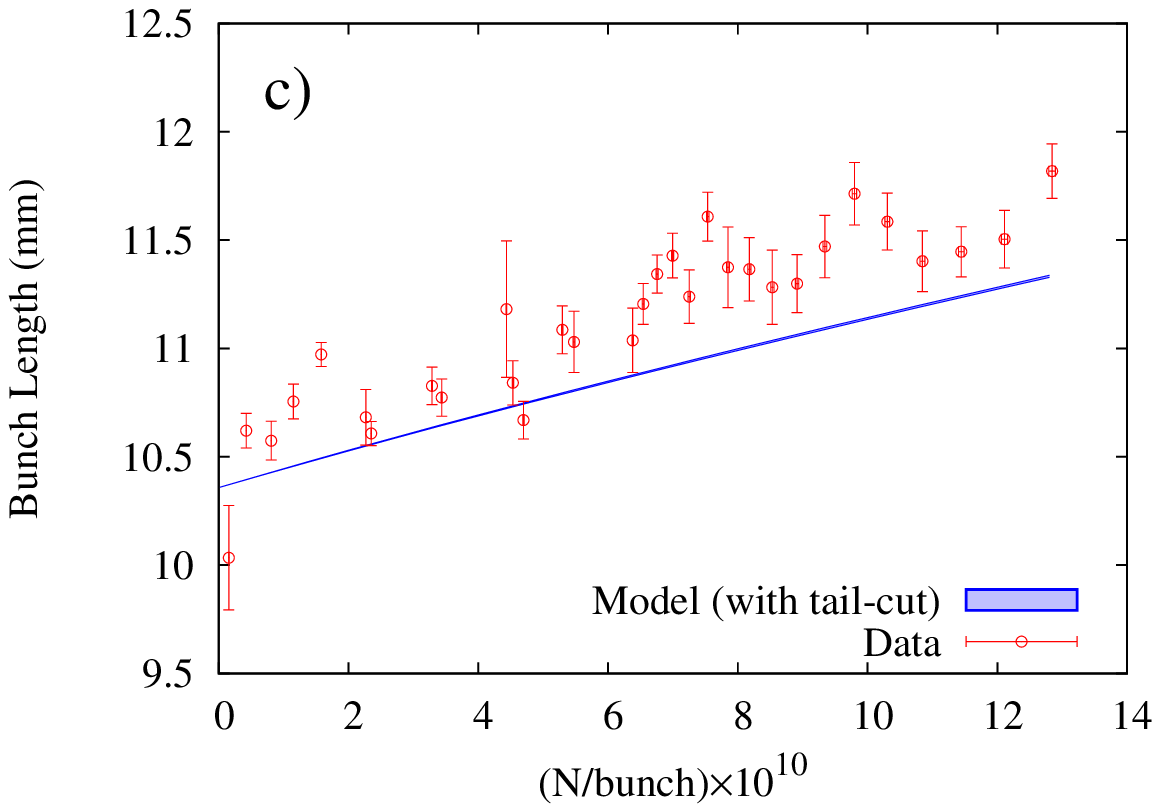}
  \label{fig:6B-Data-long}
}
\caption{\protect\subref{fig:6B-Data-horiz} Horizontal, 
\protect\subref{fig:6B-Data-vert} vertical,
and \protect\subref{fig:6B-Data-long} longitudinal 
beam size versus current for $e^-$ bunch 
with increased zero current vertical emittance.
\label{fig:6B-Data}}
\end{figure}

Data from an e$^-$ run where the vertical emittance was increased are shown
in Fig.~\ref{fig:6B-Data}.
The horizontal emittance is $4.2$ nm-rad at zero 
current and $5.5$ nm-rad at $4.8\times10^{10}$ particles/bunch.
The vertical emittances that bound the data 
are $172$~pm and $188$~pm.

\begin{figure}
\centering
\subfloat{
  \includegraphics[width=0.90\columnwidth]{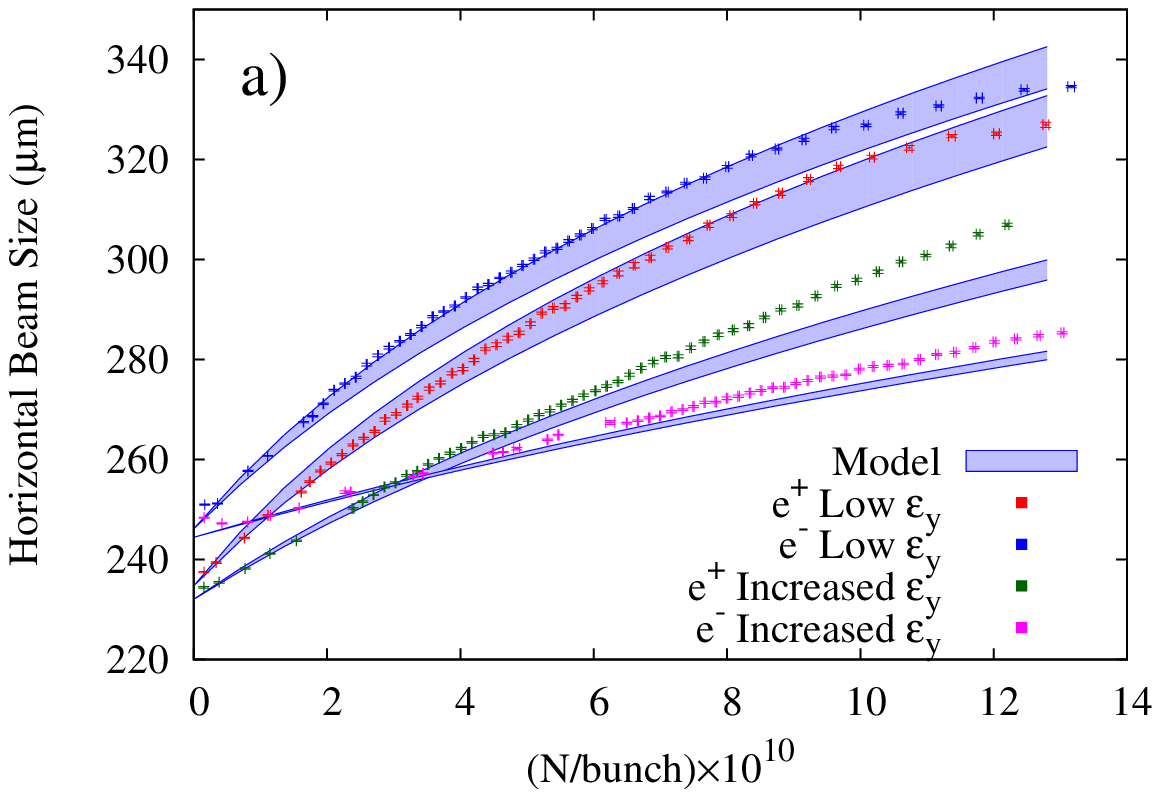}
  \label{fig:agg-Data-h}
}\\
\subfloat{
  \includegraphics[width=0.90\columnwidth]{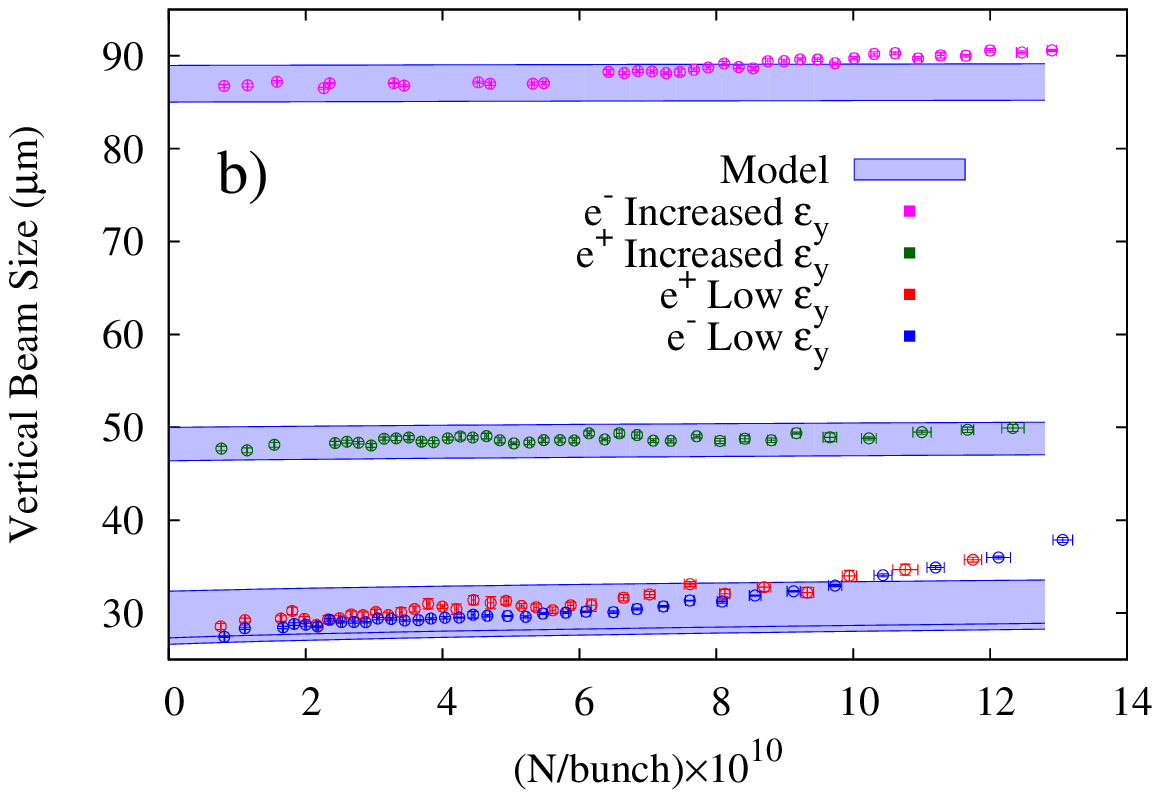}
  \label{fig:agg-Data-v}
}\\
\subfloat{
  \includegraphics[width=0.90\columnwidth]{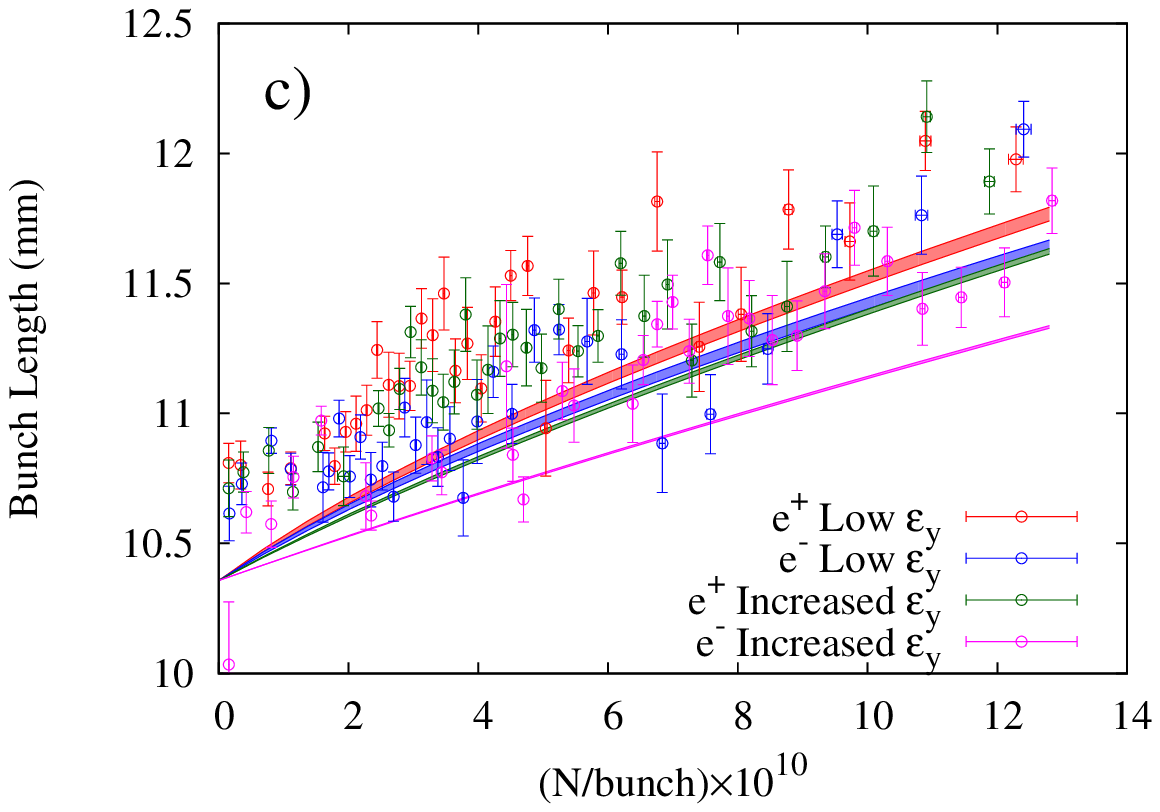}
  \label{fig:agg-Data-l}
}
\caption{Aggregated \protect\subref{fig:agg-Data-h} 
Horizontal, \protect\subref{fig:agg-Data-v} vertical,
and \protect\subref{fig:agg-Data-l} longitudinal 
data comparing $e^+$ and $e^-$ in minimum emittance 
conditions and conditions where the zero current vertical
emittance was blown up using closed coupling and dispersion bumps.
\label{fig:agg-Data}}
\end{figure}

Figure \ref{fig:agg-Data} shows the combined data 
from the two $e^-$ and two $e^+$ data sets.

\section{Discussion}
\subsection{Data}
IBS effects are most evident in the horizontal dimension, 
where large horizontal dispersion leads to significant blow-up.
In comparison, IBS is not a strong effect in the vertical.
This is because the vertical dispersion is so small.  The
direct transfer of momentum from the horizontal to the 
vertical by IBS is small at high energy.

The amount of the blow-up can be 
controlled by varying the vertical emittance, and thus the particle density.
The simulations show bunch lengthening due to IBS, but we are unable to 
distinguish IBS lengthening from potential well 
distortion in our measurements.

An interesting anomaly we have encountered is 
the behavior of the vertical beam size at high currents.  The effect is 
seen in Figs.~\ref{fig:4B-Methods-vert} and \ref{fig:68-Data-vert} 
above $8\times10^{10}$ particles/bunch.
We observe that vertical beam size plotted versus current increases 
with positive curvature.  Much more severe cases of this blow-up have 
been observed during the machine studies.  We find that adjusting betatron 
and synchrotron tunes during experiments affects the blow-up, 
but in a somewhat unpredictable way.
The blow-up is observed in both electron and positron beams.

The observed blow-up in the vertical does not appear to be an 
instrumentation effect because when the vertical size blows up,
the horizontal size drops.  This is because the blow-up in the vertical
reduces the particle density, which reduces the IBS effect in the
horizontal.

At high current, the vertical beam centroid position over $32768$ turns 
was recorded using the xBSM.
A fast Fourier transform of this data does not show a clear signal above 
background, so we cannot attribute the anomalous growth in vertical size to 
an instability.  Adjustments to the corrected chromaticity did not impact
the blow-up.

\begin{figure}[h]
\begin{center}
\includegraphics[width=1.00\columnwidth]{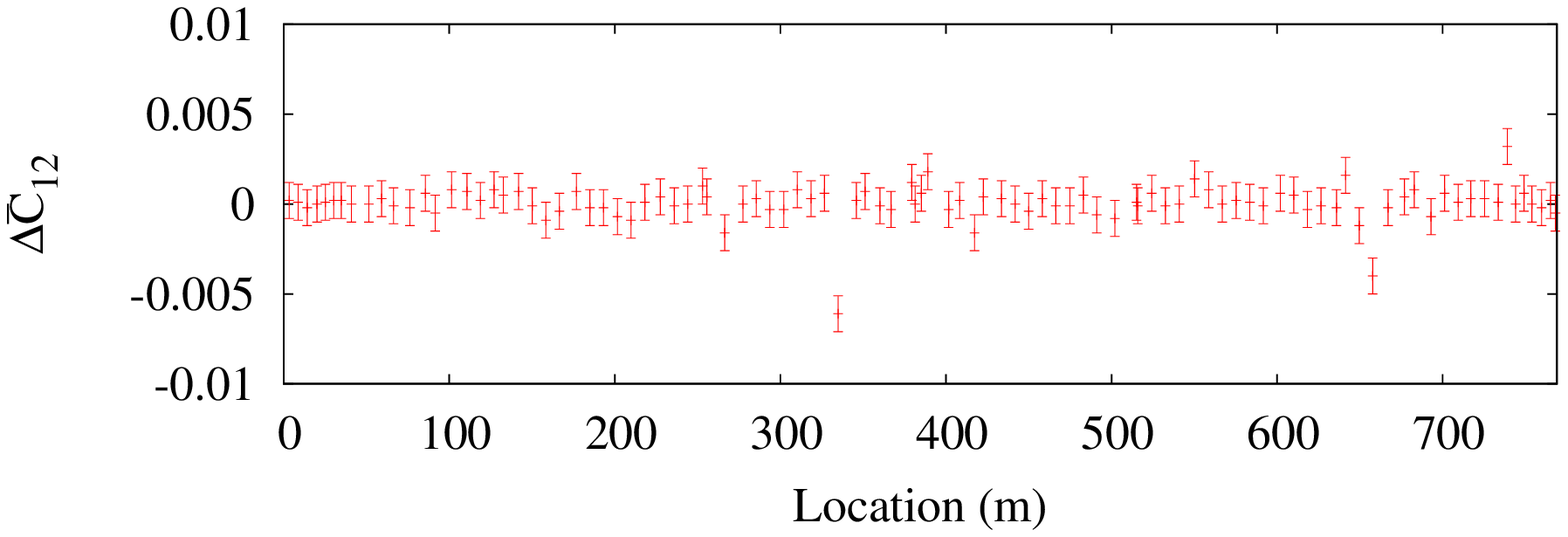}\\
\includegraphics[width=1.00\columnwidth]{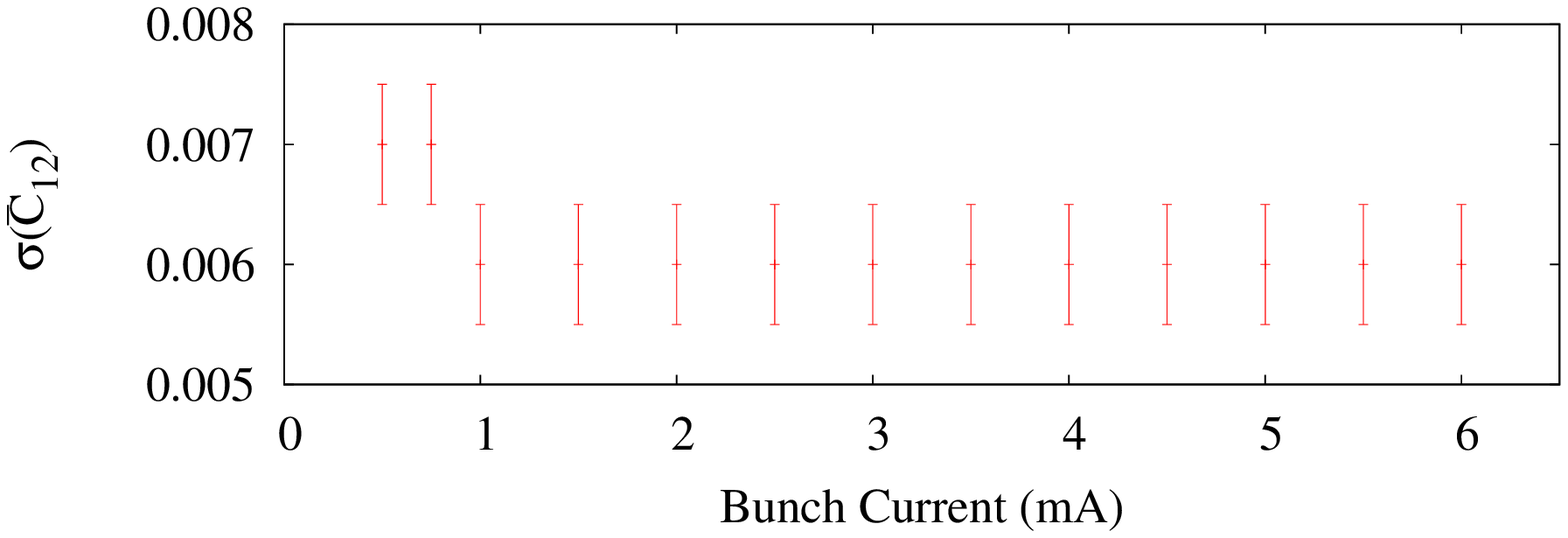}\\
\includegraphics[width=1.00\columnwidth]{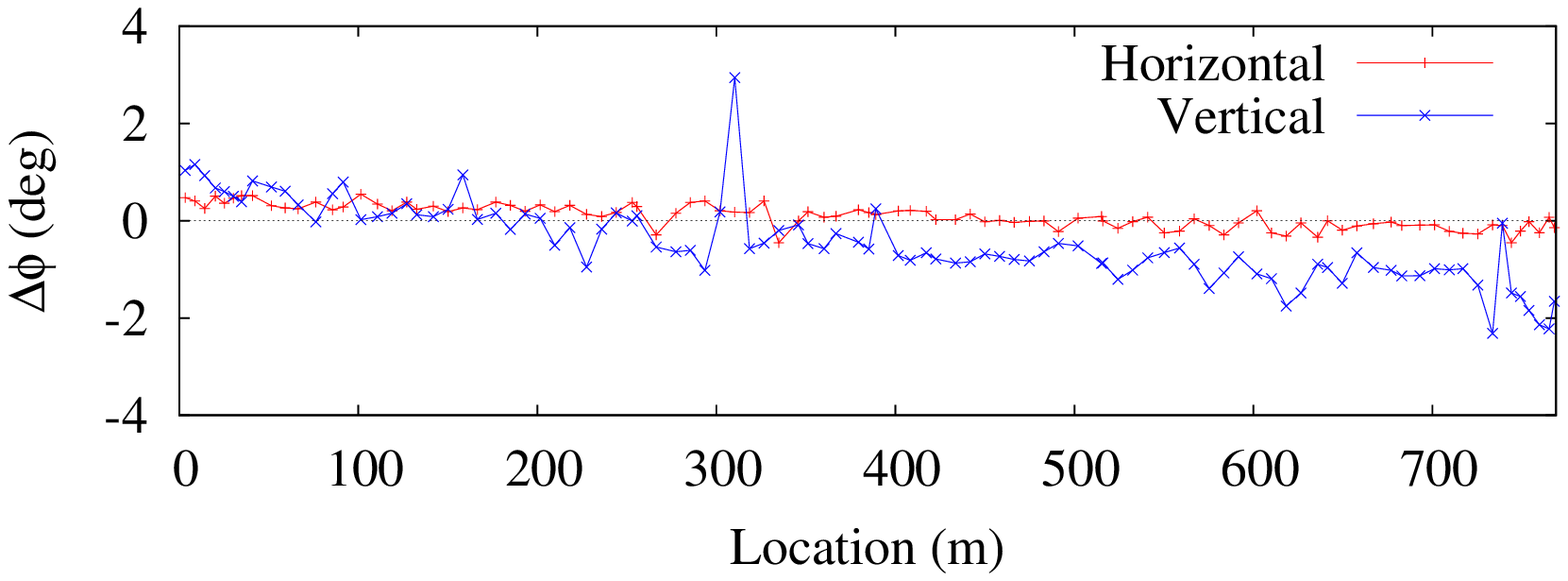}
\caption{Top: Difference in transverse coupling at $6$~mA vs $0.8$~mA
  single bunch current measured at each BPM. 
  Middle: Current dependence of rms $\left(\sigma\right)$ of residual coupling of 
  all $N$ BPMs where 
  $\sigma^2=\frac{1}{N}\sum_{i=1}^{N=100}\left(\bar{C_i}_{12}\right)^2$.
  Bottom: Difference in betatron phase at $6$~mA vs.~$0.8$~mA single bunch
  current.  The downward trend in the vertical corresponds to the current
  dependent tune shift.  $\beta$ is related to the phase advance from one
  BPM to the next.  The fractional change in $\beta$ from $6$~mA to $0.8$~mA
  is less than $1$\%.}
\label{fig:coupling_vs_current}
\end{center}
\end{figure}

Measurements of coupling $\bar{C}_{12}$ at different bunch currents are shown 
in Fig.~\ref{fig:coupling_vs_current}. There is no evidence of significant
current-dependent transverse coupling.

The measured bunch length shown here is consistently about $0.5$~mm 
longer than the predicted value.  Measurement results from 
December 2012 do not show
this discrepancy.  Between April and December 2012 the streak camera
was realigned and the analysis software was improved.  However, the evidence
does not point to any particular instrumentation systematic.

\subsection{Theory}
The presence of the Coulomb log is a well-known ambiguity in IBS theory as 
it requires the introduction of loosely defined cutoffs in the 
minimum and maximum scattering angle.  The choice of one event per
damping time as the boundary between multiple-event and single-event 
scattering is somewhat arbitrary.  That said, the data shown here are in 
reasonable agreement with theory, suggesting that with implementation of the 
tail cut, the IBS theory is a reasonable model of performance for electron
and positron machines.
Furthermore, as shown in Fig.~\ref{fig:tail-cut-rates:b}, the theory gives 
a good description of the data even when the large angle cutoff used 
in the calculation is varied by more than an order of magnitude.

The theory used here is Kubo and Oide's $\mathbf{\Sigma}$-matrix based IBS 
formalism. 
This model is a generalization of Bjorken and Mtingwa's formalism that can 
handle arbitrary coupling of the horizontal, vertical, and longitudinal 
motion.  It includes the tail-cut.  Coupling in CesrTA for the experiments
shown here was not large enough to noticeably impact the IBS growth
rates.  If coupling were significantly larger, then the predictions
from Kubo and Oide's method may diverge from those of Bjorken and
Mtingwa's method.  Additional IBS studies with more strongly coupled beams
are planned to investigate this regime.

\section{Conclusion}
We have presented data on intrabeam scattering in a high-intensity, 
wiggler-dominated, $e^-$/$e^+$ storage ring.
Additional current-dependent effects, such as tune shift and potential 
well distortion, have been observed in the beams.  An anomalous blow-up in 
the vertical dimension is seen at high current and requires further study.

This data has been compared to a generalized version of the Bjorken-Mtingwa 
formalism for calculating IBS effects.  The results presented here suggest 
that, provided the tail-cut procedure is applied, existing IBS theory 
is an accurate predictor of machine performance in $e^+$/$e^-$ machines.

Further IBS studies at CesrTA include adjusting wiggler parameters to 
observe how the equilibrium emittance of an IBS-dominated beam depends on 
the damping time, measurements at $1.8$, $2.3$, and $2.5$~GeV, and 
measurements with different RF voltages.  
There are also plans to explore IBS in strongly coupled beams.
Some of these studies have
been completed and are documented in \cite{ipac-ibs:2012} 
and \cite{ehrlichman:2013}.

\begin{acknowledgments}
The authors would like to thank Mike Billing for his many helpful
discussions.  We would like to thank Avi Chatterjee for his assistance
during the December 2012 experiments.  The experiments reported here
would not have been possible without the diligent support of the
CESR Operations Group.

This research was supported by 
NSF and DOE contracts PHY-0734867, PHY-1002467,
PHYS-1068662, DE-FC02-08ER41538, DE-SC0006505, and the Japan/US
Cooperation Program.
\end{acknowledgments}

\bibliography{prst_ibs}

\end{document}